\documentclass[twocolumn,times]{aastex63}

\usepackage{amsmath,amstext}
\usepackage[T1]{fontenc}
\usepackage{apjfonts} 
\usepackage[figure,figure*]{hypcap}
\usepackage[normalem]{ulem}             

\usepackage{graphicx}
\usepackage{color}
\usepackage[normalem]{ulem}

\newcommand{\rsun}{\mbox{\,$\rm R_{\odot}$}}        
\newcommand{\Dthree}{D$_3$}

\submitjournal{ApJ}

\shorttitle{Coronal Helium}
\shortauthors{Del Zanna, G. et al.}

\begin{document}

\title{Helium line emissivities in the solar corona}

\correspondingauthor{G. Del Zanna}
\email{gd232@cam.ac.uk}

\author{G. Del Zanna}
\affiliation{DAMTP, Centre for Mathematical Sciences, University of Cambridge, \\
  Wilberforce Road, Cambridge CB3 0WA, UK}

\author{P. J. Storey}
\affiliation{Department of Physics and Astronomy, University College London, \\
  Gower Street, London, WC1E 6BT, UK}


\author{N. R. Badnell}
\affiliation{Department of Physics, University of Strathclyde, \\
  Glasgow, G4 0NG, UK}

\author{V. Andretta}
\affiliation{INAF, Osservatorio Astronomico di Capodimonte, \\
  Salita Moiariello 16, 80131 Naples, Italy}




\begin{abstract}
We present new collisional-radiative models (CRMs) for helium in the quiescent solar corona, and predict the emissivities of the He and He$^+$ lines to be observed by DKIST, Solar Orbiter, and Proba-3. We discuss in detail the rates we selected for these models, highlighting several shortcomings we have found in previous work. As no previous complete and self-consistent coronal CRM for helium existed, we have benchmarked our largest model at a density of 10$^{6}$~cm$^{-3}$ and temperature of 20,000~K against recent CRMs developed for photoionised nebulae. We then present results for the outer solar corona, using new dielectronic recombination rates we have calculated, which increase the abundance of neutral helium by about a factor of two. We also find that all the optical triplet \ion{He}{1} lines, and in particular the well known \ion{He}{1} 10830 and 5876~\AA\ lines are strongly affected by both photo-excitation and photo-ionisation from the disk radiation, and that extensive CRM models are required to obtain correct estimates. Close to the Sun, at an electron density of 10$^{8}$ cm$^{-3}$ and temperature of 1~MK, we predict the emissivity of the \ion{He}{1} 10830~\AA\ to be comparable to that of the strong \ion{Fe}{13} coronal line at 10798~\AA. However, we expect the \ion{He}{1} emissivity to sharply fall in the outer corona, with respect to \ion{Fe}{13}. We confirm that the He$^+$ Lyman $\alpha$ at 304~\AA\ is also significantly affected by photo-excitation and is expected to be detectable as a strong coronal line up to several solar radii.
\end{abstract}



\section{Introduction}
\label{sec:intro}

Neutral and ionised helium produce the most important
transitions in the solar atmosphere after neutral  hydrogen.
Despite decades of observations and models, the helium spectrum 
is still not well understood.
Most of the attention in the literature has focused on explaining
the strong emission formed in the transition region
between the chromosphere and the corona. The main problem there
is that observed radiances in the quiet Sun are typically an
order of magnitude greater than most models predict.
The literature is very extensive. 
Interested readers are referred to e.g. the discussions
in \cite{andretta_etal:2003} and references therein or, more recently but limited to the EUV helium spectrum, in \cite{golding_etal:2017}.
Modelling the helium emission is complex, as
radiation transfer, photo-ionization,
time-dependent ionization and many other
physical effects are at play there.

In this paper, we focus on the helium spectrum produced
in the low-density (10$^7$--10$^8$ cm$^{-3}$)
hot (about 1 MK) outer solar corona, because of two main aspects.
Firstly,  several upcoming facilities will soon routinely observe
helium lines in the corona. Secondly, as far as we are aware,
no complete models of the helium emission in the corona have been
previously developed that included both neutral and ionised helium and considering all the relevant atomic processes relevant for the solar corona. 

It may at first seem surprising that transitions from neutral or singly ionised ions could be observed and that needed to be modelled in $\sim$1~MK plasmas.  It should however be noted that the strongest coronal transition is the Ly$\alpha$ from neutral hydrogen, despite the fact that the number density of hydrogen atoms is about 10$^{-7}$ compared to protons at coronal temperatures. However,  \cite{gabriel:1971} with a simple model based on the atomic rates available at the time showed that indeed this low abundance is enough to explain the radiance of this line, which is mainly formed by resonant photo-excitation (PE) from  the Ly$\alpha$ disk emission, the strongest UV spectral line. Similar calculations show that transitions from He in corona could indeed be observed.

The strongest coronal transition from helium is the doublet
Ly$\alpha$ from \ion{He}{2}, at 303.8~\AA.
The best direct measurements of this line were obtained with 
CHASE (Coronal Helium Abundance Experiment)  on board Spacelab 2
\citep{patchett_etal:1981,breeveld_etal:1988}.
On-disk and off-limb measurements of the \ion{H}{1} and \ion{He}{2}
Ly$\alpha$ lines were used by \cite{gabriel_etal:1995} to provide the
first estimate of the helium abundance at 1.15\rsun.
This  turned out to be 7.9\%, close to the photospheric value.

Routine broad-band observations from SoHO EIT
described by \cite{delaboudiniere:1999}
indicated that the \ion{He}{2} 303.8~\AA\ line is visible out to large distances,
and the author suggested that this was due to this line 
being also significantly resonantly scattered.
This is probably the case, but results from  broad-band imaging
are questionable, as a significant  contribution from
stray light and coronal lines is normally present. In particular,
the strong resonance line from \ion{Si}{11} at 303.3~\AA\ is a
significant contributor, as discussed in \cite{andretta_etal:2012},
where models of the coronal emission were discussed. 

The \ion{He}{2} 303.8~\AA\  transition, together with the
\ion{H}{1} Ly$\alpha$,  has also been observed with the
Helium Resonance Scattering in the Corona and HELiosphere (HERSCHEL) 
sounding rocket in 2009, as described recently by
\cite{moses_etal:2020}. Peculiar abundance patterns have been noted.
The  \ion{He}{2} 303.8~\AA\  line is of particular interest for the near future as one of
the remote-sensing instruments on board Solar Orbiter, the
Extreme Ultraviolet Imager (EUI, see \citealt{rochus_etal:2020})
will routinely observe the solar corona with a passband
centred on this line. The novelty is that the  full-Sun imager
has a very wide field-of-view (FOV), about 
3.8$^o$ $\times$ 3.8$^o$, corresponding to a radial FOV of
14.3\rsun\ at 1 AU and 4.0\rsun\ at perihelion, around 0.3 AU.
The FOV will therefore overlap with that of the 
Metis coronagraph \citep{antonucci_etal:2020}, which produces
narrow-band images in the \ion{H}{1} Ly$\alpha$.
By combining EUI and Metis observations, we could study the
helium abundance variations, using the present He model.  

It should be mentioned that another \ion{He}{2} line, the Balmer $\gamma$ multiplet at 1084.9~\AA, has also been observed in the corona, for instance by the SoHO spectrographs SUMER \citep{wilhelm_etal:1995} and UVCS \citep{kohl_etal:1995}.  In particular, some of the very few estimates of the helium abundance in the corona have been based on measurements of that line \citep{raymond_etal:1997,laming_feldman:2001,laming_feldman:2003}. In this paper we will not discuss further the formation of this line in the corona, although it might be observable by the upcoming Solar-C\_EUVST mission \citep{shimizu_etal:2019}.

\begin{figure}[htbp!]
\begin{center}
\includegraphics[angle=0,trim=10 20 10 30,width=\linewidth]{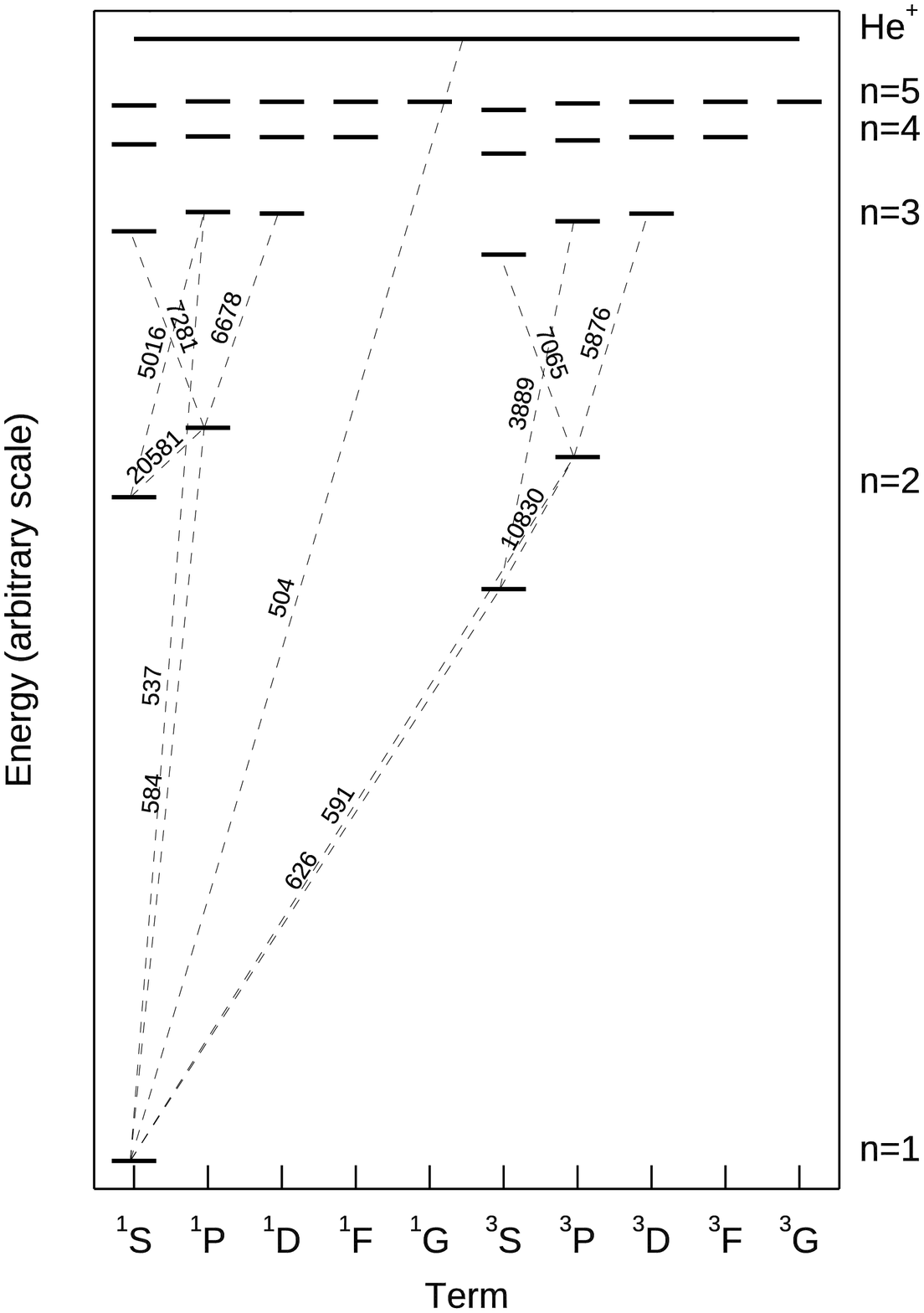}
\end{center}
\caption{A Grotrian diagram for \ion{He}{1} up to $n=5$. The main transitions discussed in this paper are shown as dashed lines, along with their wavelengths in \AA\ (air wavelengths for $\lambda>2000$ \AA).}
\label{fig:he_grotrian}
\end{figure}

Regarding neutral helium, two lines (see Figure~\ref{fig:he_grotrian})
are particularly important. The near-infrared (NIR) 10830~\AA\ and the
\Dthree\ line at 5876~\AA.  The first is one of the main transitions
which are going to be observed routinely with the  
Daniel K. Inouye Solar Telescope (DKIST, see \citealt{rimmele_etal:2015}) 
CryoNIRSP spectropolarimeter (PI: Jeff Kuhn, see \citealt{fehlmann_etal:2016} for details).
CryoNIRSP  will carry out routine daily 
observations up to 1.5\rsun\  in selected wavelength regions
between about 5000~\AA\ and 5~$\mu$m,
with a spectral  resolution of 30,000, a spatial resolution of about 1\arcsec, 
and a field of view (FOV) of 4\arcmin\ $\times$3\arcmin.
The 10830~\AA\  line is very close to the two \ion{Fe}{13}  forbidden lines at 10747 and 10798 \AA\ (air wavelengths),
the strongest NIR coronal lines (for a discussion of the coronal lines
available in the visible and NIR see \cite{delzanna_deluca:2018}).

The \Dthree\ line  will be observed routinely by the large
coronagraph on Proba-3, ASPIICS
(Association of Spacecraft for Polarimetric and Imaging Investigation of the Corona of the Sun,
see e.g. \citealt{renotte_etal:2016}.  ASPIICS will provide polarimetry
in a narrow-band centred on the \Dthree\ line, together with one on the green
\ion{Fe}{14} line. 
Both lines have been observed in the corona, the \Dthree\
only in prominences, whilst the 10830~\AA\ line sometimes also in the corona.  In both cases, there is an exciting prospect: to use the polarimetric observations to measure the coronal magnetic field.
However, their formation mechanism is unclear, in particular for the
10830~\AA\ line, as different explanations have been put forward.

Ground-based observations during eclipses generally did not indicate a coronal emission in the 10830~\AA\ line (see e.g. \citealt{penn_etal:1994}), which is always strong when prominences are present. 
However, \cite{kuhn_etal:1996} reported eclipse measurements where the  10830~\AA\ line has a brightness comparable to those of 
 the \ion{Fe}{13} NIR forbidden lines,
out to some radial distances. This was later confirmed by other ground-based observations, see e.g
\cite{kuhn_etal:2007,moise_etal:2010,judge_etal:2019}.
\cite{kuhn_etal:2007} pointed out that the width of the line was much narrower than the coronal \ion{Fe}{13} NIR forbidden lines, indicating that the emission in this line is not coronal in origin.
The He I  emission did not appear to be correlated with and hence due to cool prominence material which is often found close to the solar limb.

Given the very low neutral helium abundance expected in the corona
(around 10$^{-9}$ of the helium abundance), because  most of the
atoms will be $\alpha$ particles or He$^+$ around 1 MK, it seemed
unlikely that the 10830~\AA\ line is coronal, so various
formation mechanisms have been discussed\footnote{see, e.g. the
  SHINE Session No. 16: On the Origin of Neutrals and Low
Charge Ions in the Corona}. Some geocoronal contribution cannot be discounted in
ground-based observations, nor can the presence of cool
emission associated with e.g. filament eruptions
\citep[see, e.g.][]{ding_habbal:2017}.
It is however worth noting that no observations of the \ion{He}{1} 584~\AA\ have been reported in association with Coronal Mass Ejections (CMEs), while the \ion{He}{2} 1085~\AA\ is often observed \citep{giordano_etal:2013}.

Neutral He emission from the \ion{He}{1} 584~\AA\ line has on the other hand been observed very far from the Sun \citep{michels_etal:2002}. Such observations have been interpreted as Sun light scattered by neutral helium that enters the solar system with the local interstellar wind \citep[see, e.g.][and references therein]{lallement_etal:2004}. However, observations at different times indicate that, while plausible for measurements at large distances ($> 8-9 \rsun$), this is not a likely explanation for near-Sun measurements \citep{moise_etal:2010}. Another possible explanation put forward in the literature is desorption of atomic helium from circumsolar dust, as described e.g. in \cite{moise_etal:2010} and references therein.

Given the lack of any coronal model of the helium emission,
the main aim of the present paper is to provide one.
Section~2 describes the atomic data and models developed, while
Section~3 presents a short benchmark of the largest of our models against earlier studies of the recombination spectrum of neutral He for low-temperatures. Section~4 then provides 
a few estimates of the emissivities of the
He lines, compared to one of the \ion{Fe}{13} NIR forbidden lines,
for a sample range of coronal quiet Sun conditions.
Estimates of the expected radiances of the helium lines for a
range of solar coronal conditions, and of emissivities for nebular astrophysics are deferred to follow-up papers.

\section{Atomic data and models}

As briefly mentioned in the introduction, a significant number of models
have been developed over the decades to model the helium emission.
Most of them studied the formation mechanisms in the transition region, i.e.
where the physical processes are quite different from those present in the
solar corona. For example, in the transition region
temperatures are much lower, processes such as dielectronic
recombination are not so relevant, and densities are much higher. 

Several collisional-radiative models (CRMs) have also been developed since the
early 1970's to describe the recombination spectrum of neutral helium for nebular astrophysics.
We did not find any of the previous models suitable for our purpose, for several reasons.
The main reason is that these models were developed to study the emission from very low-temperature (from hundreds to a few tens of thousands of K) and low
density (electron density less than approximately $10^6$/cm$^3$) plasmas which are ionized by photons. As a result, dielectronic recombination, the main recombination process at coronal temperatures, was not included. However, a comparison with our largest model at an intermediate density (10$^{6}$ cm$^{-3}$) is feasible and is presented in Section~3, as a benchmark.

For our models, we have taken a `brute-force' approach, i.e. we have
developed a CRM where we include all the important levels 
for all the charge states, build a large matrix with all the rates
connecting the levels, and then obtain, assuming steady state,
the populations of all the levels at once. Once we obtained the populations, we could then calculate the $b_{nl}$ departure coefficients, check that the higher levels were statistically redistributed in $l$, that the $b_{nl}$ join smoothly to the $b_n$ and that the $b_n$ have the correct behaviour as $n \rightarrow \infty$
We had initially developed such approach to study the
various physical effects (e.g. photo-ionization, density) on 
the charge states of carbon in the  transition region, as described
in \cite{dufresne_delzanna:2019}. We  have written most of the
codes used here  in IDL, to take advantage of some of the rates and functionalities
present in the CHIANTI package, in particular those developed by one of us
(GDZ) for version 9 \citep{dere_etal:2019}, where a simplified
two-ion CRM was developed to calculate the intensities of the
satellite lines.

Our approach differs from previous ones, where e.g. hydrogenic
approximations were used, or levels were not $LS$ resolved, or  
Rydberg levels were `bundled' \citep[see, e.g.][]{burgess_summers:1969}.
We first created matrices with all the main rates affecting the
bound levels in neutral He, which is the most complex one and
took a very long time to build. 
We then created the matrices for He$^+$, using
the available CHIANTI model. We then combined them,
adding one level for He$^{++}$, and all the main level-resolved
ionisation/recombination rates connecting the three atoms.
We have then added photo-excitation (PE) and photo-ionisation (PI)
from the solar disk.

In  sophisticated CRM such as 
 COLRAD for H-like ions, see \cite{ljepojevic_etal:1984,colrad},
the lower levels are $J$-resolved.
Higher $LS$-resolved and $n$-resolved levels are then added.
In order to simplify the modelling, we have chosen for
neutral He models
where we consider only  $LS$-resolved and $n$-resolved levels.
We have kept the $J$-resolved CHIANTI model for  He$^+$. 
Our initial model had  all $LS$-resolved He levels up to $n=10$,
then $n$-resolved levels up to $n=300$.
We call this model $n=10$ CRM.  The upper bound state 
was chosen as most dielectronic recombination from He$^+$ goes to
levels below $n=300$. 
We have subsequently created more extended models,
with $LS$-resolved He levels up to $n=20, 30, 40$,
and still  $n$-resolved levels up to $n=300$.
This was done to tackle different density regimes and to
check convergence of results. 
Just to give an idea of the complexity of such models,
the larger one has 1639 $LS$-resolved levels for neutral He, for a total CRM with 1899 levels. 
In what follows, we provide a
relatively short description and discussion of the main
rates used in our models.

\subsection{He energies}

For the $LS$ levels up to $n=10$ we have used the  
 ab-initio calculations of \cite{drake_morton:2007}, which are regarded
 as more accurate than the observed ones.
 For the higher levels,
we have used the updated coefficients presented in
\cite{drake:2006} to calculate the quantum defects and the energies 
for all the levels with $l \le 7$. The levels
with higher angular momentum $l$ are basically degenerate, so we have
set their energies equal to that of the $l = 7$ levels, for each $n$.


\subsection{He A-values }

\cite{drake_morton:2007} provided A-values in LS coupling
for most of the transitions up to $n=10$.
For the transitions involving higher levels we initially calculated values using
{\sc autostructure} \citep{badnell:2011}. 
We used standard methods to obtain the scaling 
parameters for the lower orbitals, assuming a Thomas-Fermi-Amaldi
central potential to obtain energies relatively close to
those from Drake.

However, after many tests, we have found that the A-values
for some transitions to high levels were not accurate.
This is a general problem, related to the fact that for higher levels the configuration expansion 
is unbalanced due to the absence of all bound states above the limit of the calculation, and continuum states. 
An improvement  is obtained by switching off
$LS$-mixing among levels.
{\sc autostructure} was also
modified (by NRB) to experiment with $LS$-mixing among
only lower levels, but this does not solve the problem of orbital term dependence for the $2^3S$ and $2^1S$.
We therefore opted to use A-values obtained using the method described by 
 \cite{bates_damgaard:1949}, which only requires the quantum defects of the initial 
 and final states.
For all the remaining transitions, we have calculated
the A-values with the hydrogenic approximation using the
code RADZ1 \citep{storey_hummer:1991}.

Finally, there are three very important rates which affect
substantially the model atom. 
For the forbidden 1s$^2$ $^1$S$_0$ -- 1s 2s $^3$S$_1$
transition we use a probability of 1.27$\times$10$^{-4}$/s from
\cite{lach_pachucki:2001}. This value is close
to the measurement by \cite{woodworth_moos:1975},  1.1$\times$10$^{-4}$/s.
We note that CHIANTI uses a value of 1.73$\times$10$^{-4}$/s.
For the two-photon 1s$^2$ $^1$S$_0$ -- 1s 2s $^1$S$_0$  transition we use 
50.94/s, the same as in the CHIANTI model, which originates
from \cite{drake:1986}.
Finally, for the important 1s$^2$ $^1$S$_0$   -- 1s 2p $^3$P 
$LS$ transition we use the value 59.2/s. We note that the
value calculated by \cite{drake_morton:2007} for the
1s2 $^1$S$_0$ -- 1s 2p $^3$P$_1$  is 177.6/s, nearly the same
as \cite{lach_pachucki:2001}, 177.5, but different than the value
used in the CHIANTI model: 233/s. 
Recent measurements by \cite{dall_etal:2008} give
177$\pm$8, in excellent agreement with the theoretical values.

The above rates are for the $LS$-resolved levels.
In order to connect the $LS$-resolved levels with the
$n$-resolved levels, we have formed a statistically weighted average of the A-values from all $l$ states of a given $n$ to each lower $LS$ state, and then for each lower
$LS$-resolved level extrapolated for any  $n$ value.
Typically, the averaged values decrease as $n^{-5}$.

In order to connect the $n$-resolved levels
we used hydrogenic values. As we have encountered problems with the
approximations recommended by \cite{burgess_summers:1976},
we have resorted to use the analytic expressions (see e.g. \citealt{goldberg:1935}) for the gf values and
the Gaunt factors pre-calculated with very good accuracy by \cite{morabito_etal:2014}. 
The A-values were then calculated using the energies derived from the quantum defects.

\subsection{He electron collisional rates among lower levels}

The most accurate electron collisional rates
were calculated by \cite{bray_etal:2000} in $LS$ coupling.
Only those from the lowest 4 levels (to all the levels up to $n=5$),  were provided. However, they are the most important rates, as levels with $n>2$ are not significantly populated. One limitation of these data is that
rates were calculated only up to log $T=5.75$.
We have therefore extended these rates by extrapolating in the
\cite{burgess_tully:1992} scaled domain, using high-energy limits. For the dipole-allowed transitions, we used the $gf$ values obtained from Drake's A-values. For the forbidden transitions, we calculated the limits
 with {\sc autostructure}. 

\begin{figure}[htbp!]
\begin{center}
\includegraphics[angle=-90, width=8.5cm]{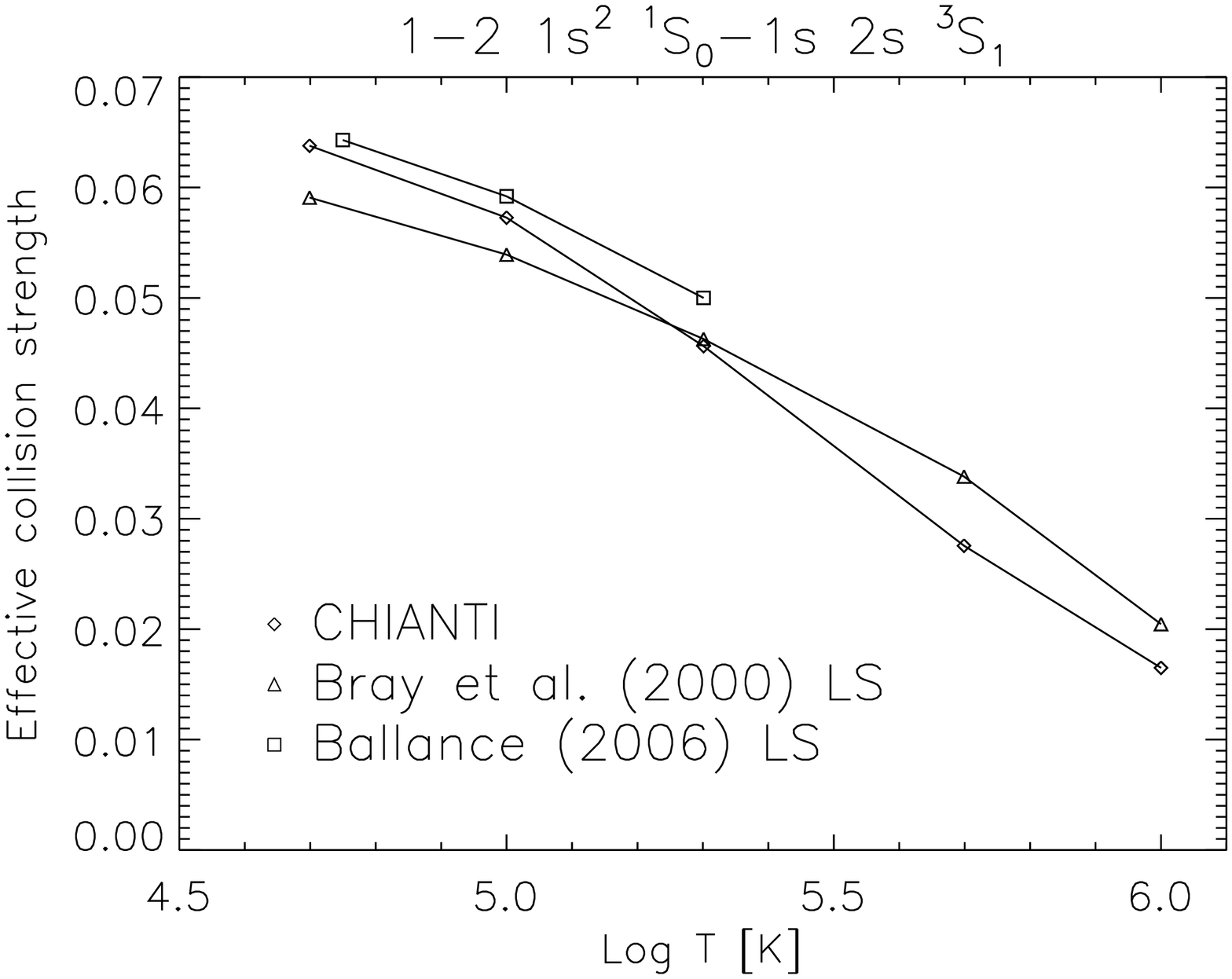}
\includegraphics[angle=-90, width=8.5cm]{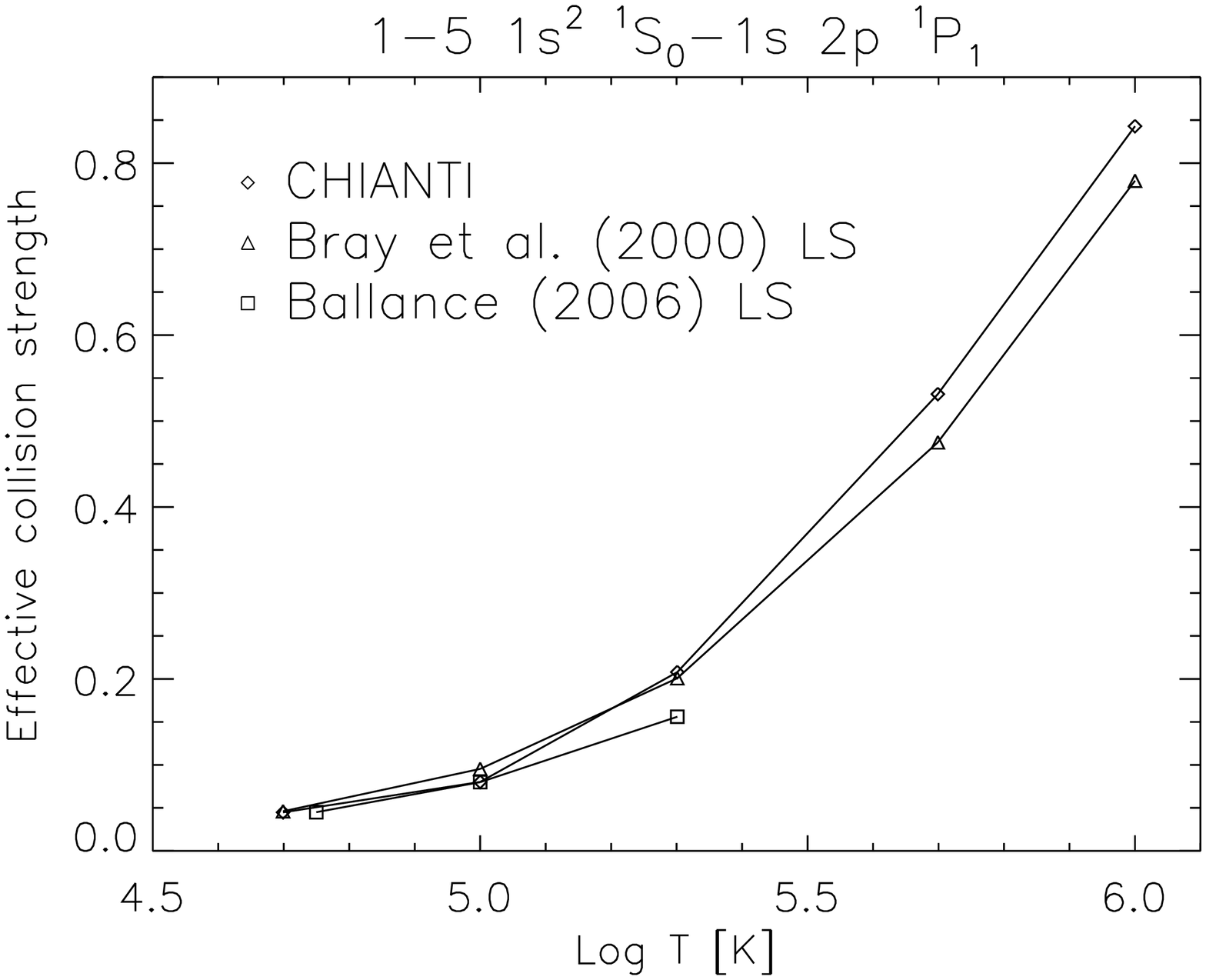}
\end{center}
\caption{The effective collision strengths for two of the main
  transitions in He, the forbidden one to the metastable and
the resonance transition. The Bray et al. values have been extrapolated.}
\label{fig:he_ups}
\end{figure}

When comparing the rates with those present in the CHIANTI
database, we found several inconsistencies, even for
transitions between singlets, see e.g. Figure~\ref{fig:he_ups}.
However, we have verified that these differences in the rates
do  not significantly affect the intensities around 1 MK.
We did this by  building  an $LS$-resolved model with all levels
up to $n=5$, to match the CHIANTI model. 
On a side note, we have also uncovered an error in the rates
affecting the $n=2 \, ^3$P, the upper level of the 10830~\AA\
transition in the photo-ionization code CLOUDY \citep{ferland_etal:2017}, which is also using the rates from \cite{bray_etal:2000}.

We have also compared the above rates with those
calculated by \cite{ballance_etal:2006}
with the $R$-matrix suite of codes, including pseudo-states.
We found general agreement, but also some differences, see e.g. Figure~\ref{fig:he_ups}.
\cite{ballance_etal:2006} only included   19 $LS$ terms up to $n=4$, and the maximum
temperature was only 2$\times 10^5$ K.

\subsection{He electron/proton collisional rates for higher levels}

  Rate coefficients for electron collisions with He were calculated with the {\sc autostructure} distorted wave (AS/DW) code. Coefficients were  calculated up to $n=40$ and compared with those obtained from the Impact Parameter (IP) approximation for dipole transitions \citep[see][]{williams:1933, alder_etal:1956, seaton:1962}, implemented with a rectilinear trajectory as described in \cite{storey_thesis}. However, significant differences were found between the two methods and this was attributed to the fact that for dipole transitions the Coulomb Bethe (CBe) approximation, which is used to top-up the contribution from high partial waves (typically >30) in the   AS/DW code, overestimates the contribution from those smaller partial waves \citep{burgess_tully78:1978} which correspond to the projectile impact parameter being less than the target orbital radius. For example, for a 7p orbital, the partial wave corresponding to the orbital radius is 162 for an energy of 10~Ryd, so that the partial waves from 30 to 162 are computed with the CBe method in a domain where it overestimates. The IP method avoids this problem by starting the integration over impact parameter at the orbital radius.  
  We therefore adopted the IP rates for all dipole allowed transitions  for all $n > 5,\,  \delta n=1,\, \delta l=1$, for both electrons and protons. Similar issues can arise with non-dipole transitions, which cannot be obtained from the IP method. However the non-dipole transitions are unimportant compared to the dipole $l$-changing collisions in redistributing population among the $l$ substates, while the collision strengths for transitions from the 2$^3$S metastable state to higher states, which play a significant role in populating those states, should be reliable.

  For the collisional rates from the populated lower states, we adopted 
  an $n^{-3}$ extrapolation of the \cite{bray_etal:2000} values from the 
  ground state, the 2s $^1$S, 2s $^3$S, and
  2p $^3$P to the $n=5$ levels. 
  To connect these populated lower states to the $n$-resolved levels, we have summed the \cite{bray_etal:2000} rates to the $n=5$ levels and adopted the same $n^{-3}$ extrapolation, up to $n=300$. The results are very close to the AS DW ones 
  for lower $n$ values. We have also experimented using the AS DW values
  instead of the extrapolated ones, and found very little differences in the main results.
  
  To connect the last $LS-$resolved levels with the
first $n$-resolved level, we have used the IP rates as described above, summing over the accessible $nl$ states.

For the transitions connecting  the $n$-resolved levels, the main process
is collisional excitation and de-excitation by electrons. 
We have coded the \cite{percival_richards:1978} approximation.
The strongest rates are those where $\Delta n=1$, however we have added
also those with $\Delta n=2,3,4$. 

We have also coded the semi-empirical rates recommended by
\cite{vriens_smeets:1980}, and for our coronal temperatures 
found very small differences (less than 10\%) with the
\cite{percival_richards:1978} rates.
We note that \cite{guzman_etal:2019} compared the
\cite{vriens_smeets:1980} rates with those calculated by other
authors, also finding good overall agreement.

\subsection{He $l$-changing collisional rates}

$l$-changing ($\Delta l=1$)  collisions are very effective in
redistributing the populations of levels within each
$n$ shell. For the transitions among the non-degenerate levels
with lower $l$ we used the IP approximation (the same as in the previous Section), and
added all the rates for electrons, protons,
He$^+$, and He$^{++}$. We note that the electron rates are
significant for low $l$, but then the proton rates become
dominant. 

\begin{figure}[htbp!]
\begin{center}
\includegraphics[angle=-90,width=8.5cm]{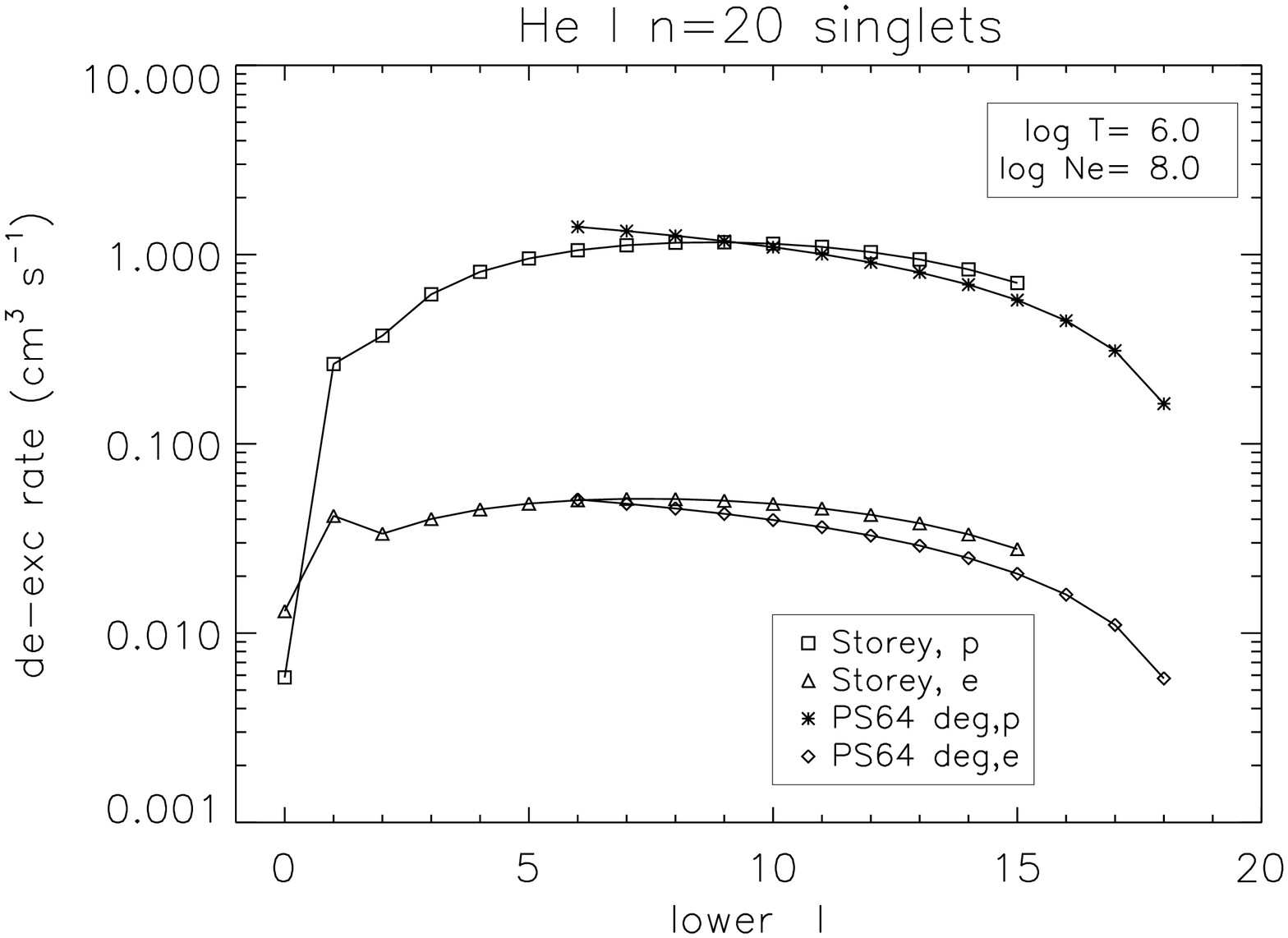}
\includegraphics[angle=-90,width=8.5cm]{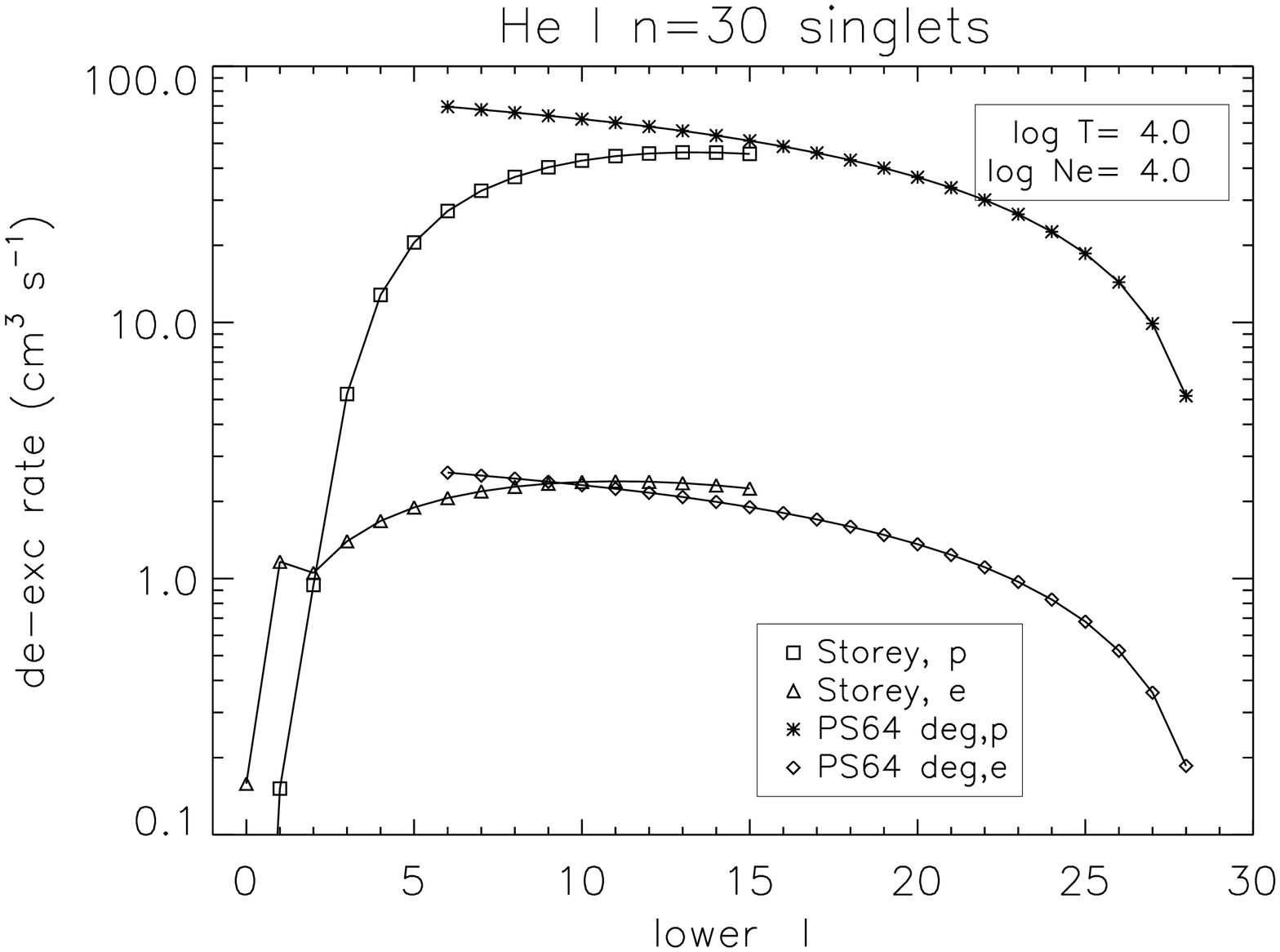}
\end{center}
\caption{Top: l-changing de-excitation rates for the $n$=20 singlets,calculated with the IP approximation (Storey) and with the classical \cite{pengelly_seaton:1964} (PS64), for electrons and protons at coronal conditions. Bottom: the same, but for $n$=30 and nebular conditions.}
\label{fig:ps64}
\end{figure}

Regarding the transitions among the degenerate levels,
we have coded the classical approximation by
\cite{pengelly_seaton:1964} for electrons, protons, He$^+$, and He$^{++}$.
For our purposes (i.e. high
temperatures), this is still a very accurate approximation,
as discussed in detail by \cite{guzman_etal:2017}.
We have visually  checked that all the IP rates for
the  non-degenerate levels converge to the degenerate values, see Fig.~\ref{fig:ps64} (top) as an example for $n=20$. 

When benchmarking our codes, we have uncovered
a few  problems in the IP rates for electrons and protons described by \cite{guzman_etal:2017}
and implemented in CLOUDY \citep{ferland_etal:2017}.
Fig.~\ref{fig:ps64} (bottom) shows the values corresponding to those displayed in their Fig.1, for electrons and protons. Their results for protons, labelled S62 in their figure, should correspond to the IP results labelled Storey, p in Fig.~\ref{fig:ps64} but are, in practice, very much smaller and do not match the values obtained from the degenerate IP method (PS64) at high $l$ as is required.
A collaborative effort is ongoing to resolve such problems.

\subsection{He collisional ionisation (CI)}

\begin{figure*}[htbp!]
\begin{center}
\includegraphics[angle=0, width=8.cm]{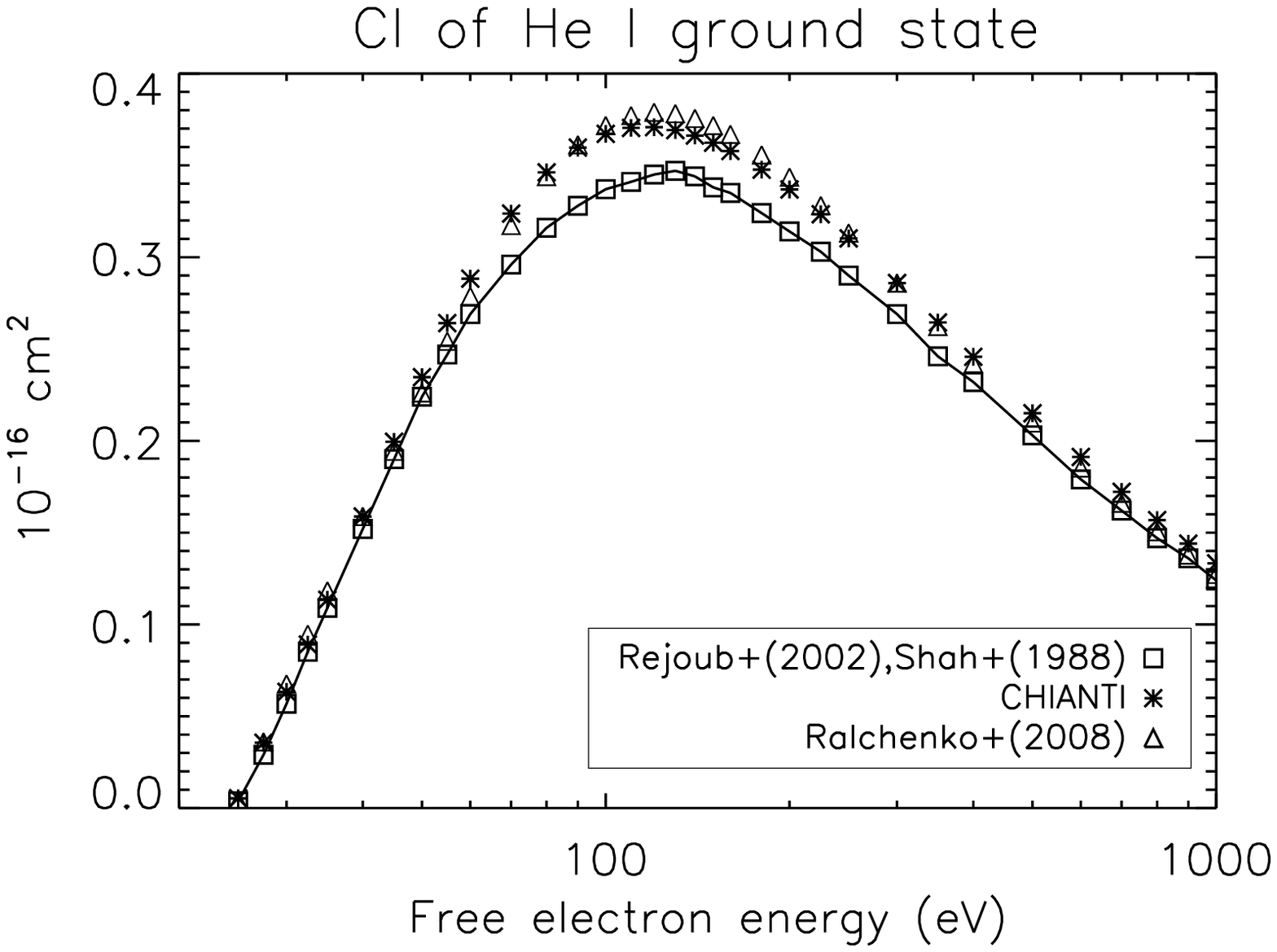}
\includegraphics[angle=0, width=8.cm]{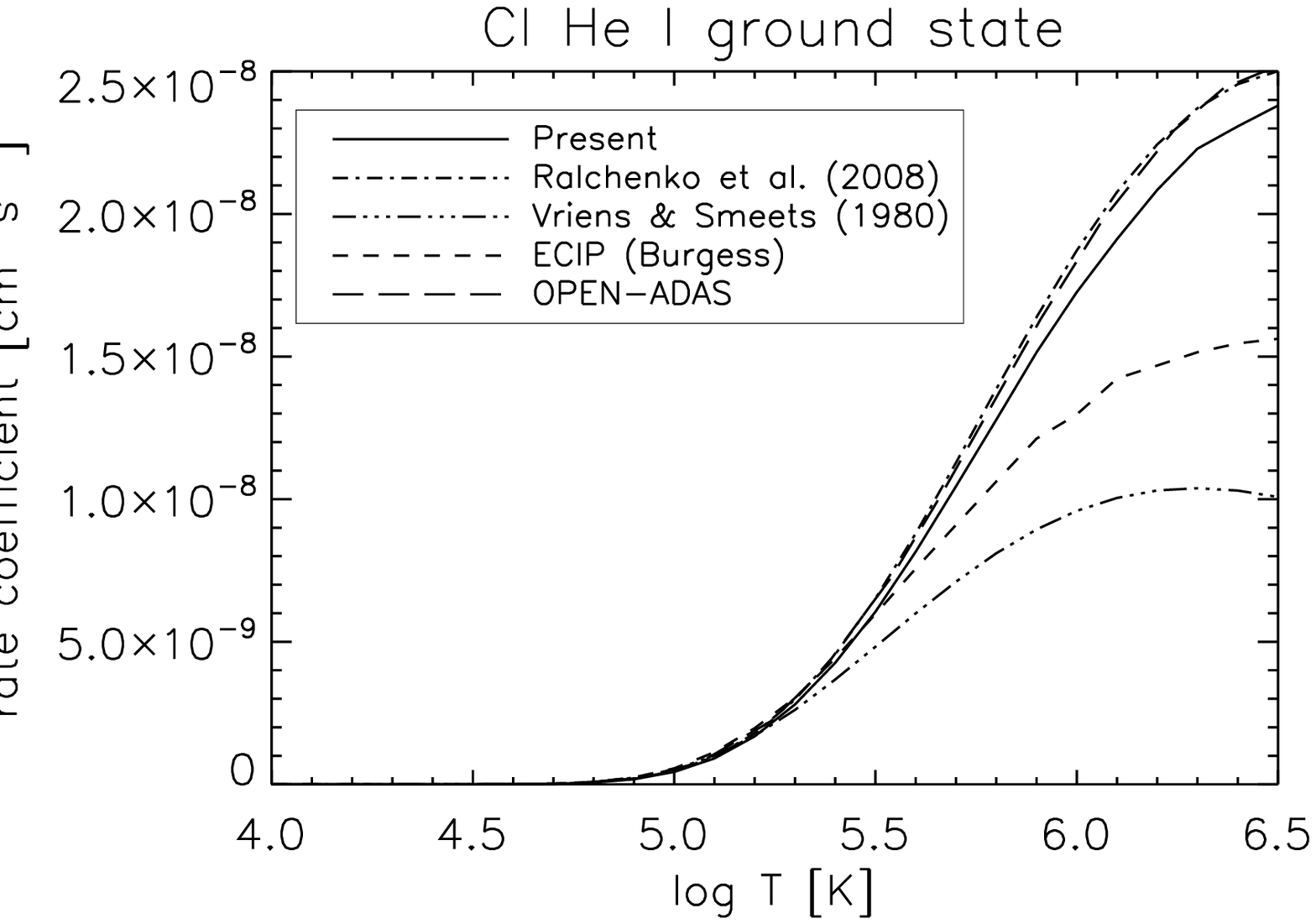}
\end{center}
\begin{center}
\includegraphics[angle=0, width=8.cm]{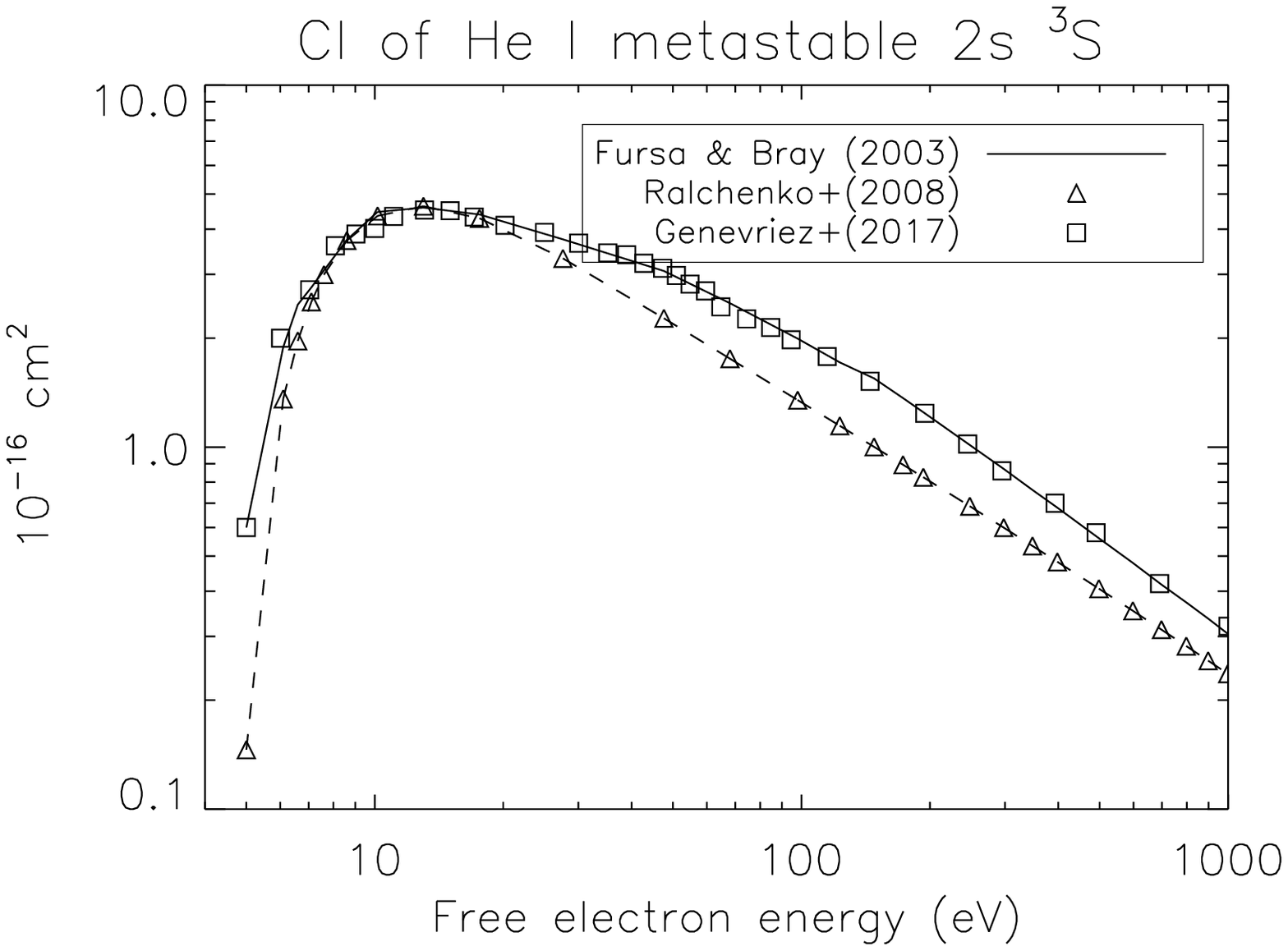}
\includegraphics[angle=0, width=8.cm]{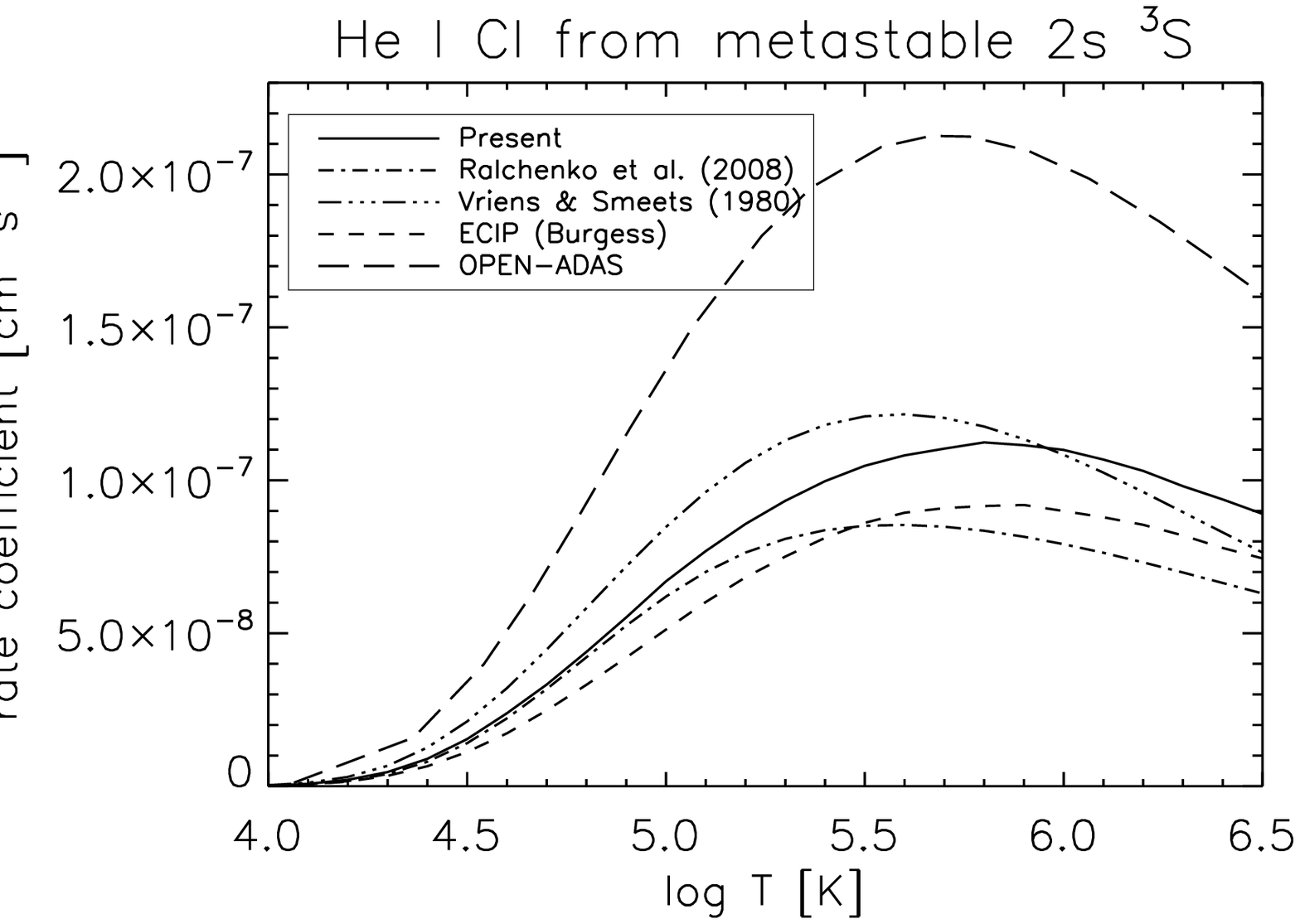}
\end{center}
\begin{center}
\includegraphics[angle=0, width=8.cm]{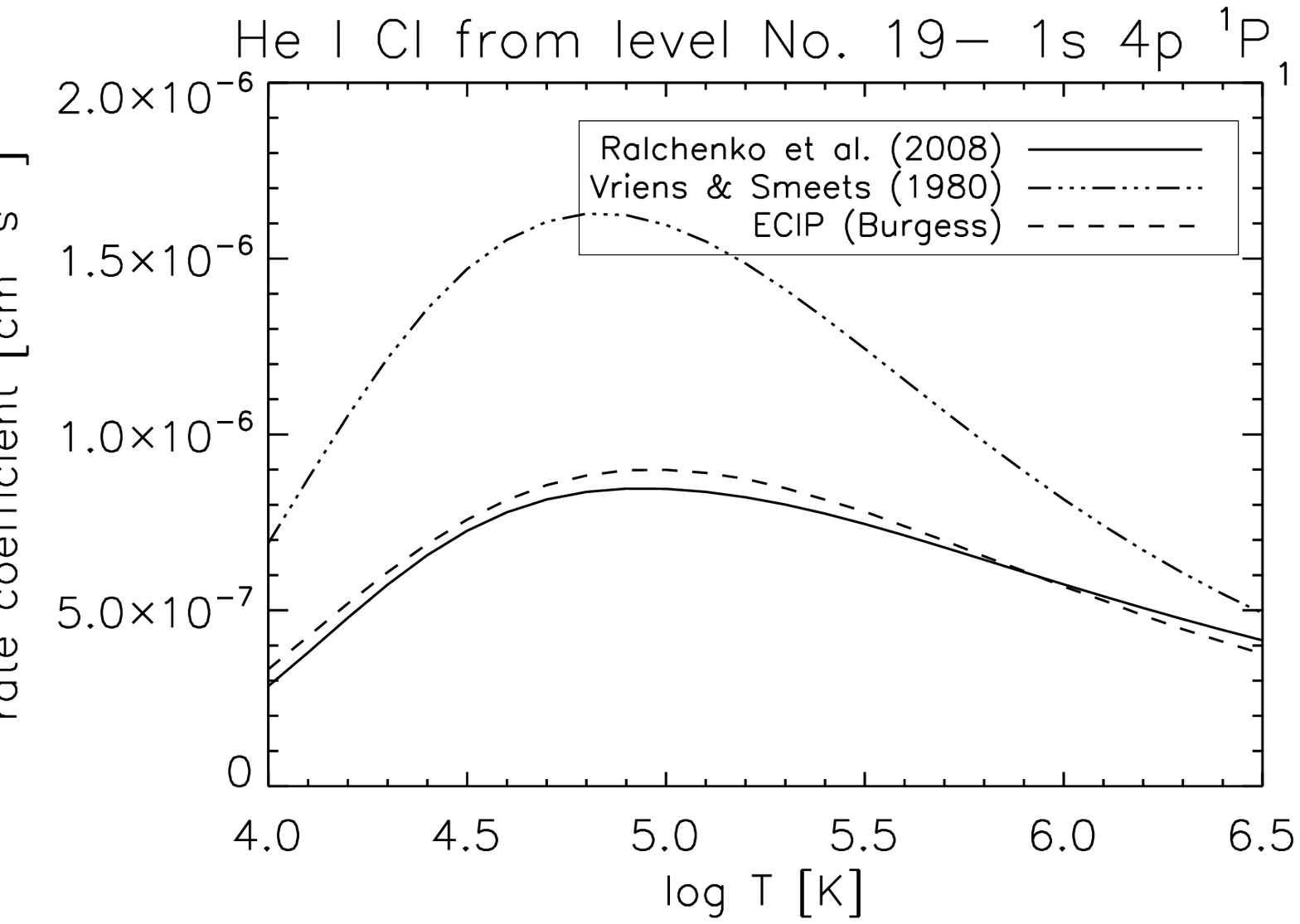}
\end{center}
\caption{Top: cross section and rate for collisional ionisation (CI) from the ground state.
  Middle and bottom plots: the cross section and rate coefficient for CI from the
  the metastable $^3$S, and the rate from the 1s 4p $^1$P.}
\label{fig:he_ci}
\end{figure*}

For the collisional ionisation (CI) from the ground state of
neutral He we have adopted the cross sections measured by
\cite{rejoub_etal:2002} and \cite{shah_etal:1988}.
As shown in Figure~\ref{fig:he_ci} (top), these values are slightly
lower than those used in the CHIANTI model to obtain the
relative abundance of neutral He.
We obtained the rate coefficients for this process
and the others discussed below with a Gauss Laguerre quadrature
\citep[see,e.g.][]{delzanna_mason:2018}.

We note that currently the CHIANTI model for ion charge states
only considers  ground states. This is common among all
astrophysical codes (e.g. CLOUDY, ATOMDB). 
However, the
population of the He metastable ($2^3$S) is significant at coronal 
densities (about 20\%), so the CI from this level is important
for the modelling. Significant discrepancies
between measurements and theory for CI from the $2^3$S were present until recent
measurements, as discussed by \cite{genevriez_etal:2017}. In their
paper, they show that the ab-initio calculations
by \cite{fursa_bray:2003} are in excellent agreement with observations.
We have therefore adopted these theoretical cross sections,
extrapolated them to high energies using classical scaling,
and obtained the rate coefficient shown in Figure~\ref{fig:he_ci} (middle).  
Following on from \cite{fursa_bray:2003}, further
cross-sections for a few of the other lower levels
(up to 1s 4p) were calculated by
\cite{ralchenko_etal:2008} who provided  fitting coefficients.
We have calculated the rates from the  excited levels
using the \cite{ralchenko_etal:2008}
fitting coefficients and our Gauss Laguerre quadrature.
Surprisingly, we found a discrepancy between the 
\cite{ralchenko_etal:2008} and the \cite{fursa_bray:2003} cross sections for
the metastable $2^3$S. The middle plot in Figure~\ref{fig:he_ci}
shows the rates.

We have then compared the rates from the other excited levels
with two semi-classical and widely used approximations which we have
coded:
the semi-empirical one of \cite{vriens_smeets:1980} and the
Exchange Classical Impact Parameter (ECIP) method developed
by A. Burgess, as described in \cite{summers_ral_report}.
For the lower levels, it is unclear which of the two approximations
is closer to the calculated values. However, for the higher levels
the ECIP one agrees much better, as shown in 
Figure~\ref{fig:he_ci} (bottom). We have therefore adopted the
ECIP rates for all the  levels  above 1s 4p in  our models.
We note that the ECIP method has the correct
behaviour at low and high energy, see e.g. \cite{burgess_summers:1976}.

Finally, for each CI rate we added to a model we also
added the rate for the inverse process,
three-body recombination, obtained assuming detailed balance.

\subsection{He$^{+}$ and He$^{++}$ charge exchange with H}

{
For the recombination process of He$^{++}$ by charge exchange with neutral
H to form  He$^{+}$, we have taken the cross sections calculated by  
\cite{zhang_etal:2010}, checked that they are consistent
with measurements at high energies \citep{havener_etal:2005},
and calculated the rates with a 12-point Gauss-Laguerre
integration. The resulting rate coefficients are significantly higher
(especially at high temperatures)  than the
rough estimate of  \cite{kingdon_ferland:1996} of 10$^{-14}$ cm$^{3}$ s$^{-1}$,
constant with temperature. For example, at 1 MK we obtained a value of
4$\times 10^{-10}$ cm$^{3}$ s$^{-1}$.  However, when taking into consideration the low
abundance of neutral H, the actual rates are much smaller than the other recombination
rates, so this effect turns out to be negligible.
The same occurs for the recombination process of He$^{+}$ by charge exchange with neutral
H (in its ground state) to form neutral He (in its ground state).
In this case we have taken the cross-sections of \cite{zygelman_etal:1989},
and extended them to higher energies by taking into account \cite{loreau_etal:2014}.
The rate coefficients for recombination of 
He$^{+}$ by charge exchange with neutral H in excited states to the
singlet and triplets in neutral He are higher \citep{loreau_etal:2018},
but as most of the population of neutral H is in the ground state, the
rates are much smaller and again negligible for our coronal conditions.

}

\subsection{He$^+$ model ion}

For the He$^+$ model ion, we have basically kept the
atomic rates present in the CHIANTI database. The model ion
includes 25 $J$-resolved levels up to $n=5$.
It uses electron collisional rates calculated by C. Ballance (priv. comm.)
with the $R$-matrix suite of codes, including  pseudostates.
The A-values are from \cite{parpia_johnson:1982}. 

\subsection{He$^+$ CI}

For the CI from the ground state of He$^+$ we have used the
CHIANTI rate, which was assessed by \cite{dere:2007}.
For the CI from excited states we have used the ECIP approximation.
We note that fractional population of the 2s $^2$S is about 10$^{-4}$
at log Ne=8, and  the rate coefficient is only a factor of 10 higher than the
rate from the ground state, so CI from the first excited level
is not as important as the CI from the metastable in He.
For each CI rate we added to a model we also
added the rate for the inverse process,
the three-body recombination, obtained assuming detailed balance.

\subsection{Radiative recombination (RR) from He$^+$  and He$^{++}$}

For the spontaneous radiative recombination (RR) from the He$^+$ ground state
we have used the
level-resolved rate coefficients calculated by NRB \citep{badnell:2006}.
As almost all the RR is to the lower levels we have included in the
models, we have just used these level-resolved rates, by
matching the ordering of the levels.

For the RR from
He$^{++}$ into He$^{+}$ we have used the total rates from \citep{badnell:2006}.

\subsection{Stimulated Radiative Recombination (SRR) from He$^+$ }

In principle, as there is a strong radiation field from the solar disk, stimulated radiative recombination (SRR) from He$^+$ might be significant. However, the ratio of the cross section for SRR to that for spontaneous RR scales like $\nu^{-3}$ where $\nu$ is the photon frequency and therefore decreases with increasing $\nu$. In practice, SRR is only significant for very high $n$ states where the ionization energies and $\nu$ are very small. We have calculated the rates for this process by taking the PI cross-sections and using the microreversibility relation to obtain the cross-sections for SSR, level-resolved. The resulting rates are small compared to the spontaneous RR rates, being about 1/4 at $n=100$ at $T=10^6$K and much less at lower $n$. As the DR rates dominate over the RR rates in any case for our coronal conditions, the SRR process can be neglected. Stimulated dielectronic recombination, SDR, is also negligible in solar coronal conditions.

\subsection{Dielectronic recombination (DR) from He$^+$}

\begin{figure*}[htbp!]
\begin{center}
\includegraphics[angle=90, width=12.5cm]{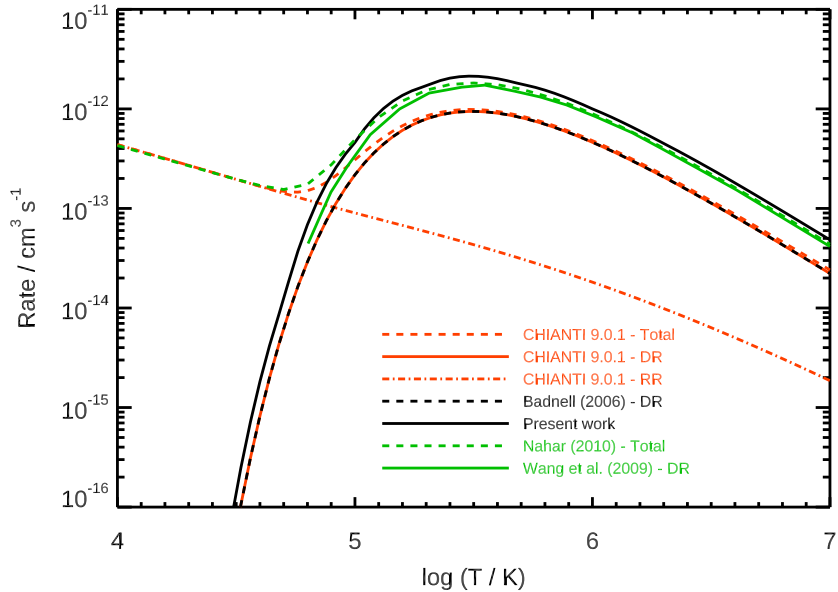}
  \caption{Total DR rate coefficients.}
\end{center}
\label{fig:total_dr}
\end{figure*}

Dielectronic recombination (DR) from He$^+$ has normally
not been included in previous CRM, as for nebular astrophysics
the dominant recombination rate is the RR. However, for our
hot coronal conditions the DR is obviously a dominant
recombination process.

We initially included in the models the level-resolved
DR rate coefficients produced by NRB as part of the DR project
\citep{badnell_etal:2003}, calculated with {\sc autostructure}
and a few post-processing tools.
These values are nearly universally used to model
laboratory and astrophysical spectra, and are included in
all the main atomic codes (e.g. CLOUDY, CHIANTI, ATOMDB).

However, when comparing the total rates with other published
literature values, we found a significant discrepancy, of about a
factor of two.  On the other hand, the results of
\cite{wang_etal:1999} and \cite{nahar:2010} were relatively close,
as was the most accurate calculation (according to A. Burgess), carried
out by J.Dubau during his PhD.
Unfortunately, none of the published calculations were usable
for our  models. For example, \cite{nahar:2010} only provided tabulated
rates for a few levels. 

The reasons for the lower DR rates obtained by \cite{badnell_etal:2003} for this ion turned out to be
related to orbital orthogonalisation.
K-shell problems are best described by non-orthogonal orbitals,
if one is using a unique orbital basis.
When the H-like sequence was originally calculated 25 years ago,
{\sc autostructure}  always
Schmidt-orthogonalized. Since then, it  has been realised that
it is better to follow Cowan's method \citep{Cowan81}, i.e. do not
orthogonalize, but just neglect the overlaps, for orbitals
that are inherently non-orthogonal. 
For He+, the overlaps are very large between the Rydberg/Continuum s,p
orbitals and the 1s, 2s, 2p. This affects the Auger rates, but the DR cross section is
not sensitive to them until n-values larger those measured 
by the original experiment and which was used to benchmark the AS calculation.

The effect drops-off very quickly along the
isoelectronic sequence with Z since the overlaps rapidly
diminish then. By C$^{5+}$, the increase in the total is barely 10\%.
We (NRB) have redone the calculations for these ions. 
The revised values for the H-like sequence can be found on
our UK APAP web site\footnote{www.apap-network.org}.

Figure~\ref{fig:total_dr} provides a comparison between
total DR rate coefficients. 
At 1 MK, the rate coefficient  as calculated by \cite{dubau_thesis} is
9.3$\times 10^{-13}$cm$^3$/s, that calculated by
\cite{nahar:2010} is 8.94$\times 10^{-13}$cm$^3$/s, while
the earlier DR project value was 4.56$\times 10^{-13}$cm$^3$/s.
The revised value with the Intermediate Coupling (IC) calculations is
1.0$\times 10^{-12}$cm$^3$/s, i.e. about a factor of two higher.
This resulted in a direct increase in the emissivities of the neutral He lines, as DR is the main recombination process in the corona.

Finally, for our models we needed to map the
$LS$-resolved and $n$-resolved rates (by interpolation in $n$) to the
level ordering in each model.
It turns out that at most about half of the total DR goes to the
$LS$-resolved  levels but to include all the DR (within a
few percent at most) we needed to include 
$n$-resolved levels  up to $n=300$ (the values are usually
calculated up to $n=999$).

\subsection{Photo-excitation (PE) for He and He$^+$}

To include photo-excitation (PE) in the model  for
He and He$^+$, we have used a modified version of the CHIANTI codes:
once the matrices with the A-values are built, the photo-excitation
and de-excitation rates are included for all the possible
transitions (calculating the wavelengths from the differences in
energies), using a dilution factor and radiation function. 
 We experimented with both a black-body of a 
temperature between 5800 and 6100 K, and observed spectra.

For the incident radiation field we have assumed
as a baseline the 1 \AA-resolution quiet Sun reference irradiance spectrum
compiled by \cite{woods_etal:2009_whi}, because it covers a
wide spectral range, from the X-rays to 2.5 microns, and is relatively
well calibrated. 
We have converted the irradiances to radiances assuming
uniform distribution on disk. 
This is  a relatively good assumption for the quiet Sun.
It neglects the limb-brightening effects but as we are not interested here
to polarization, they can be ignored. The radiances obtained in this way compare well
with on-disk radiances measured by e.g. the SoHO spectrometers.
Clearly, at UV/EUV wavelengths the spectrum can differ significantly
from that of the quiet Sun. This will be explored in a follow-up paper.

As the local ions have significant thermal velocities
in the corona, the radiation they experience does not have the fine structure (Fraunhofer lines in absorption) of the solar spectrum emitted from the disk. 
We have therefore applied a smoothing to the solar spectrum, and extended it with that of a black body 
at 6,100 K for the infrared wavelengths, between 2.5 and 50 microns. 

For the important transitions, we have used the following disk radiances:
for the 10830~\AA\ line 902681 ergs cm$^{-2}$ s$^{-1}$ sr$^{-1}$, obtained from SORCE SIM measurements;  
for the 584~\AA\ line  500.5 ergs cm$^{-2}$ s$^{-1}$ sr$^{-1}$,
obtained from SoHO CDS measurements of the quiet Sun, while for the He$^+$ 303.8~\AA\ line we have used the
quiet Sun value  of 4800 ergs cm$^{-2}$ s$^{-1}$ sr$^{-1}$ \citep[see][and references therein]{delzanna_andretta:2015}. 
We have analysed the SORCE SIM data and 
have found that solar variability does not significantly
affect the radiance at  10830~\AA. 
On the other hand, \cite{andretta_delzanna:2014, delzanna_andretta:2015} have shown that
radiances and irradiances of the EUV lines can vary substantially with the 
solar cycle (the irradiances of the He and He+ lines vary by about a factor of two). 
Therefore, further modeling will be needed to study the coronal emission when the Sun is active.

Finally, we note that the photoexcitation of the strong EUV/UV  lines depends significantly on the local coronal Doppler velocity, as comparatively little continuum emission is present at those wavelengths. 
Therefore, there is the diagnostic potential for measuring outflow velocity via Doppler dimming/brightening effects in the \ion{He}{2} 304~\AA\ line, as discussed in a follow-up paper.
This is the same principle that has been used extensively in the literature
to measure outflows using the  \ion{H}{1} Ly-$\alpha$,
following \cite{kohl_withbroe:1982,noci_etal:1987}.

\subsection{Photo-ionization (PI) for He and He$^+$}

Photo-ionization (PI) is an important process which is not
currently included in CHIANTI.  The fits to the
photoionisation cross-sections provided by \cite{verner_etal:1996}
are widely used in the literature in most photoionisation codes
(e.g. CLOUDY), however they are only the total cross-sections from the
ground states. They were obtained  by fitting the Opacity Project
(OP, see \citealt{seaton_etal:1994,seaton:2005}) cross-sections, by removing the resonances
and adjusting the thresholds to observation.
Clearly, for our models, photoionisation from excited states is
important.
As PI is the inverse process of RR, photoionisation
cross-sections from the lower levels have been calculated
with {\sc autostructure} \citep{badnell:2006}.
We have verified that these three sets of cross-sections agree for the 
ground state.


For the PI from higher levels we have used the semi-classical Kramers
hydrogenic formula in the form:
\begin{displaymath}
  \sigma(\lambda;n_i) = 8.68896\,10^{-37} \: E_i^2 \: n_i^{-2} \: \lambda^3 \: ,
\end{displaymath}
where the cross-section $\sigma$ is in cm$^2$, $\lambda$ is the ionizing
photon wavelength in \AA, and $E_i$ are $n_i$ are respectively the ionization
threshold and the principal quantum number of level $i$.  For levels with
$n\le 5$ we have also applied the Gaunt factors from
\cite{karzas_latter:1961}.
In the application
of the above formula to He$^0$ we have used the observed ionization energies,
verifying that the results reasonably match the detailed OP calculations, when
available.

We have also experimented in using the OP values, whenever available, by
adjusting the thresholds. The end results do not change significantly, 
but we opted for the semi-classical cross-sections as the effects due to the 
incorrect location and low resolution of the resonances in the OP data are difficult to quantify.

For the incident radiation field we have assumed 
the same solar spectrum used for PE.

\section{Benchmark at low-temperature}

\begin{table}[!ht]
\begin{center}
   \caption{Emissivities  (10$^{-26}$ erg cm$^{3}$ s$^{-1}$ ) of the
 strongest He lines, for $T{\rm e}$=20,000 K and  $N_{\rm e}$=10$^{6}$ cm$^{-3}$.
\label{tab:table}
}
\begin{tabular}{rllllllllllll}
\tableline
\tableline
 $\lambda$ (\AA) & levels
               &  P05 (B) & P12 (B)& F (A) & F (B) \\
\tableline
\noalign{\smallskip}
584  & S 2p-1s &         &        &  162  &       \\
2945 & T 5p-2s &  2.96   & 2.22   &  2.16 & 2.16  \\
3188 & T 4p-2s &  6.51   & 5.09   &  5.0  & 5.0  \\
3889 & T 3p-2s &  18.3   & 15.0   & 14.9  & 14.9   \\
3965 & S 4p-2s &  1.54   & 1.21   & 0.034 & 0.99    \\

4026 & T 5d-2p &   3.18  & 2.01   & 1.94  & 1.94  \\
4388 & S 5d-2p &   0.83  & 0.57   & 0.42  & 0.44  \\
4471 & T 4d-2p &   6.80  & 4.71   & 4.40  & 4.40  \\

4713 & T 4s-2p  &   0.92  & 1.05  & 1.06  & 1.06  \\
4922 & S  4d-2p &  1.80  & 1.31   & 0.90  & 0.93  \\
5016 & S  3p-2s &  4.04  & 3.17   & 0.06  & 2.64  \\
5876 & T  3d-2p &  20.2  & 18.3   & 15.7  & 15.7   \\
6678 & S  3d-2p &  5.22  & 4.88   & 2.64  & 2.75  \\

7065 & T 3s-2p  &   7.17  & 7.62   & 7.65  & 7.67  \\
7281 & S 3s-2p  &  1.36  & 1.30    & 0.88  & 1.18  \\
10830& T 2p-2s  &  255    & 215    & 213   & 214  \\
18685& T 4f-3d  & 2.37   & 0.85    & 1.15  & 1.15  \\
20587 & S 2p-2s &  -     & 5.98  & 5.1$\times$10$^{-3}$& 5.0 \\
\noalign{\smallskip}
\tableline
\tableline
\end{tabular}
\tablecomments{The first column gives the wavelength (in air, except the first and last ones in vacuum),
the second  indicates if a line is
between singlets (S) or triplets (T) and gives the transition.
         P05 (B): Porter et al. (2005) case B;
         P12 (B): Porter et al. (2012) case B;                              ;
            F (A): full n=40 model case A;  F (B): full n=40 model case B.
}
\end{center}
\end{table}

As previously mentioned, we found it useful to benchmark our largest
$n=40$ model at low temperatures against previous sophisticated CRM. 
We have chosen for the benchmark $T$=20,000 K and an electron density $N$=10$^{6}$ cm$^{-3}$, allowed by our  $n$=40 model.
The most complex CRM for neutral He 
were developed for nebular astrophysics, see for example
\cite{brocklehurst:1972,smits:1991,smits:1996,benjamin_etal:1999,porter_etal:2005,porter_etal:2012}
for neutral He and \cite{hummer_storey:1987} for ionised He.
The basic atomic rates have changed over the years, and each model was different. 
The approach in the earlier studies was often to solve the statistical balance equations in terms of departure coefficients $b$ from Saha-Boltzmann level populations. The calculations 
usually considered only $n$-resolved levels to start with, then the $b_{nl}$ for the terms (in  LS coupling) were calculated, assuming that higher levels were statistically redistributed in $l$. Matrix condensation techniques were often employed to reduce the number of levels.
Collisional ionization from the metastable level of He (or higher ones) was usually not included.
Photo-excitation and photo-ionization from anything other than the ground state was also generally not included. 

The latest two CRM on He are from \cite{porter_etal:2005,porter_etal:2012}, 
and are based on codes
available within the CLOUDY photoionisation code \citep{ferland_etal:2017}, although we note that several improvements on atomic rates (especially on the l-changing collisions) have been (and are still) ongoing in CLOUDY. 
We have chosen these two models not just because they are more recent, but especially because they used the same CE as we do in our model, from \cite{bray_etal:2000}. In fact, some differences in the line emissivities are known to be caused by the use of different CE rates.

As the main published results are line emissivities of the 
pure He recombination spectrum, we had to apply a few modifications to our coronal model. The first one is to remove 
He$^{++}$ and apply a black-body photo-ionizing spectrum with a dilution of 1.74 $\times$ 10$^{-11}$ and $T$=100,000 K to obtain well over 99\% of the population in He$^{+}$, and simulate the effects of a hot star driving a He recombination spectrum. 
The second one is to remove the DR (as normally not included in previous models, being negligible at low temperatures), and use improved RR rates obtained from the state-of-the-art photo-ionization cross-sections by \cite{hummer_storey:1998}. We used $LS$-resolved RR rates up
to $n$=40, then $n$-resolved rates up to $n$=99.

The emissivities are usually presented in the literature for case A and case B.
Case A is simply assuming an optically thin plasma. 
Most literature values are in case B, which is an approximation to model the real plasma emission when the strongest singlet lines, decaying to the ground state, are re-absorbed. To obtain the case B, we have set to zero the RR to the ground state and all the A-values of the singlet (above $n$=2) decays to the ground state.

The emissivities are 
\begin{equation}
E = \frac{4\, \pi \, j}{ N_{\rm e} \, N(He^+) } =  
\frac{h\, \nu_{ji} \, N_j \, A_{ji} \,  N(He)}
{N_{\rm e} \, N(He^+) } 
\quad {\rm erg \, cm}^3 \, {\rm s}^{-1}
\end{equation}
\noindent 
where the first definition is how the emissivities are usually
indicated in the literature, and the second one is how we calculated them: 
$N_j$ is the  population of the upper level $j$, relative to the total population of
He, N(He), i.e. He$^{0}$ and He$^{+}$; $A_{ji}$ is the radiative transition rate,
and $h\, \nu_{ji}$ is the energy of the photon.

The resulting emissivities for all the strongest He lines in the visible/near infrared are shown in Table~\ref{tab:table}.
There is generally a good agreement with the latest 
published results by \cite{porter_etal:2012}, which indicates that the present model is correct.  
However, significant differences are present with  
\cite{porter_etal:2005}, and also with earlier studies (not shown in the Table). We note that we did not expect to see such differences between the various models, as one would expect accuracy of a few percent, but defer
to a follow-up paper a full discussion on the reasons for such discrepancies.

We experimented with different rates for collisional excitation to the higher levels, as well
as different rates between the higher levels (l-changing collisions), but our 
results were unchanged. We also noticed that collisional ionization does not have a large effect at these low densities. 
The main populating processes for the upper levels of the lines shown in the Table are the RR from He$^{+}$ and collisional excitation from the metastable states.

\section{Results for coronal plasma }

\subsection{\ion{He}{1} vs. \ion{Fe}{13}}

\begin{figure*}[htbp!]
\begin{center}
\includegraphics[angle=0,trim=60 20  0 30,width=0.48\linewidth]{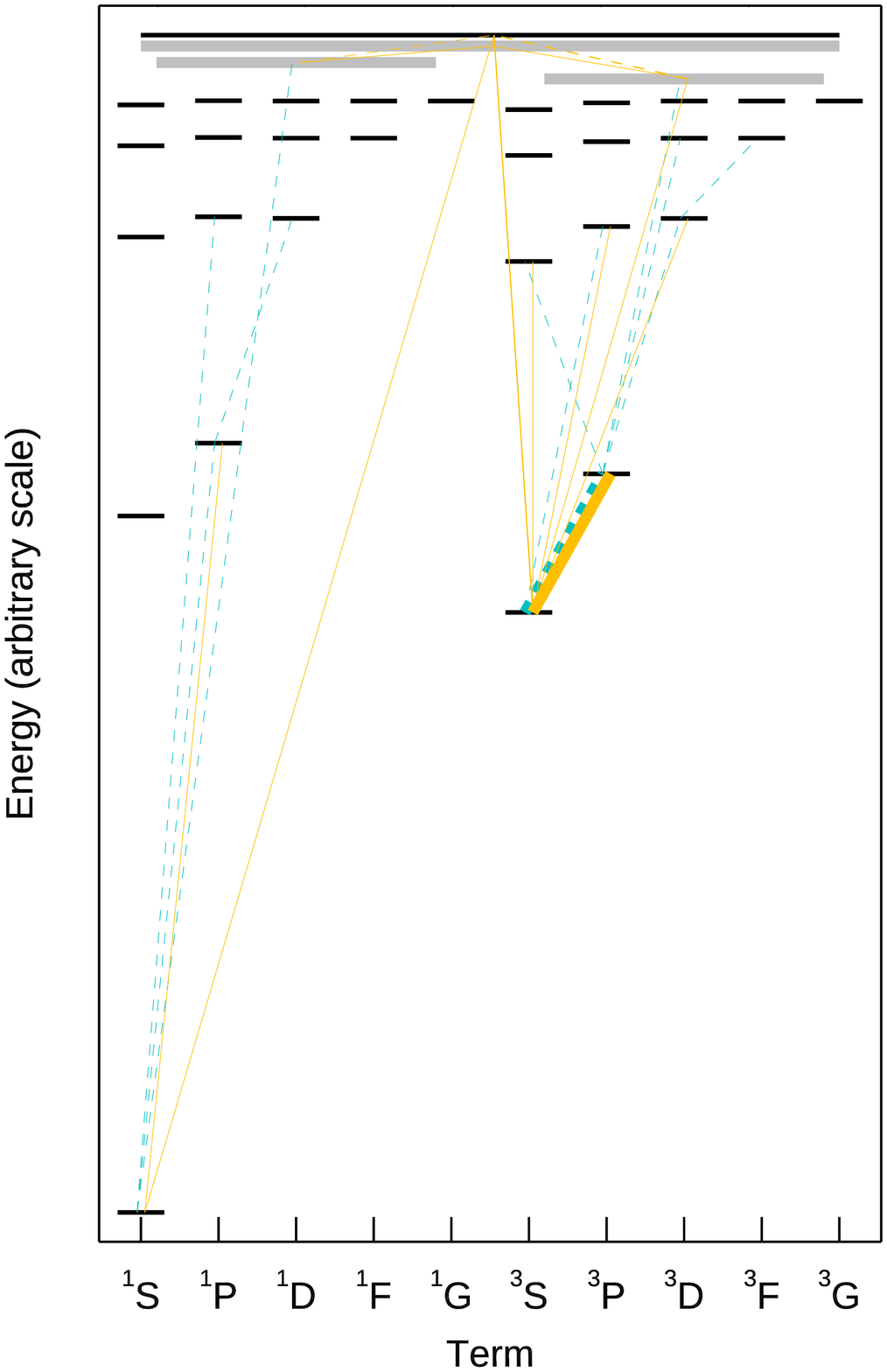}
\includegraphics[angle=0,trim= 0 20 60 30,width=0.48\linewidth]{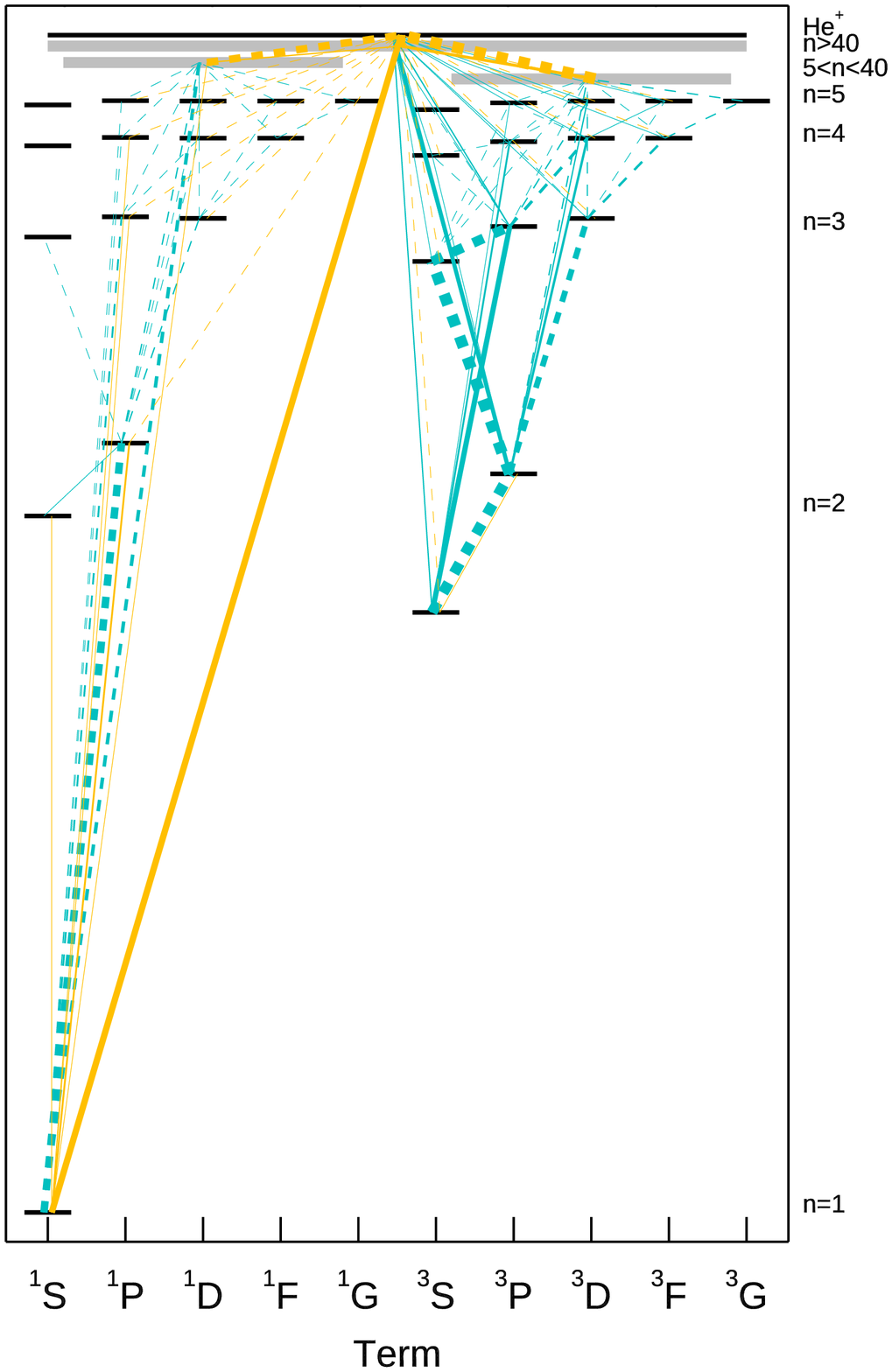}
\end{center}
\caption{The main transition rates for \ion{He}{1}. The strongest net brackets, $n_i P_{ij} - n_j P_{ji}$, are shown: orange lines represent collisional net brackets (including radiative recombinations), azure represent net radiative brackets (photoexcitations/ionisations, spontaneous/stimulated transitions). Upward and downward net rates are represented by dashed and solid lines, respectively. Only net brackets whose absolute value is greater than 1/100$^{th}$ of the largest net bracket are shown. Left and right panels show net rates in the case of zero and non-zero external radiation field, respectively, computed at 1.05~\rsun. In both cases, the electron density and temperature are $\log n_\mathrm{e}=8$ and $\log T=6$, respectively.}
\label{fig:he_net_rates}
\end{figure*}

\begin{figure}[htbp!]
\begin{center}
\includegraphics[angle=-90,width=9.cm]{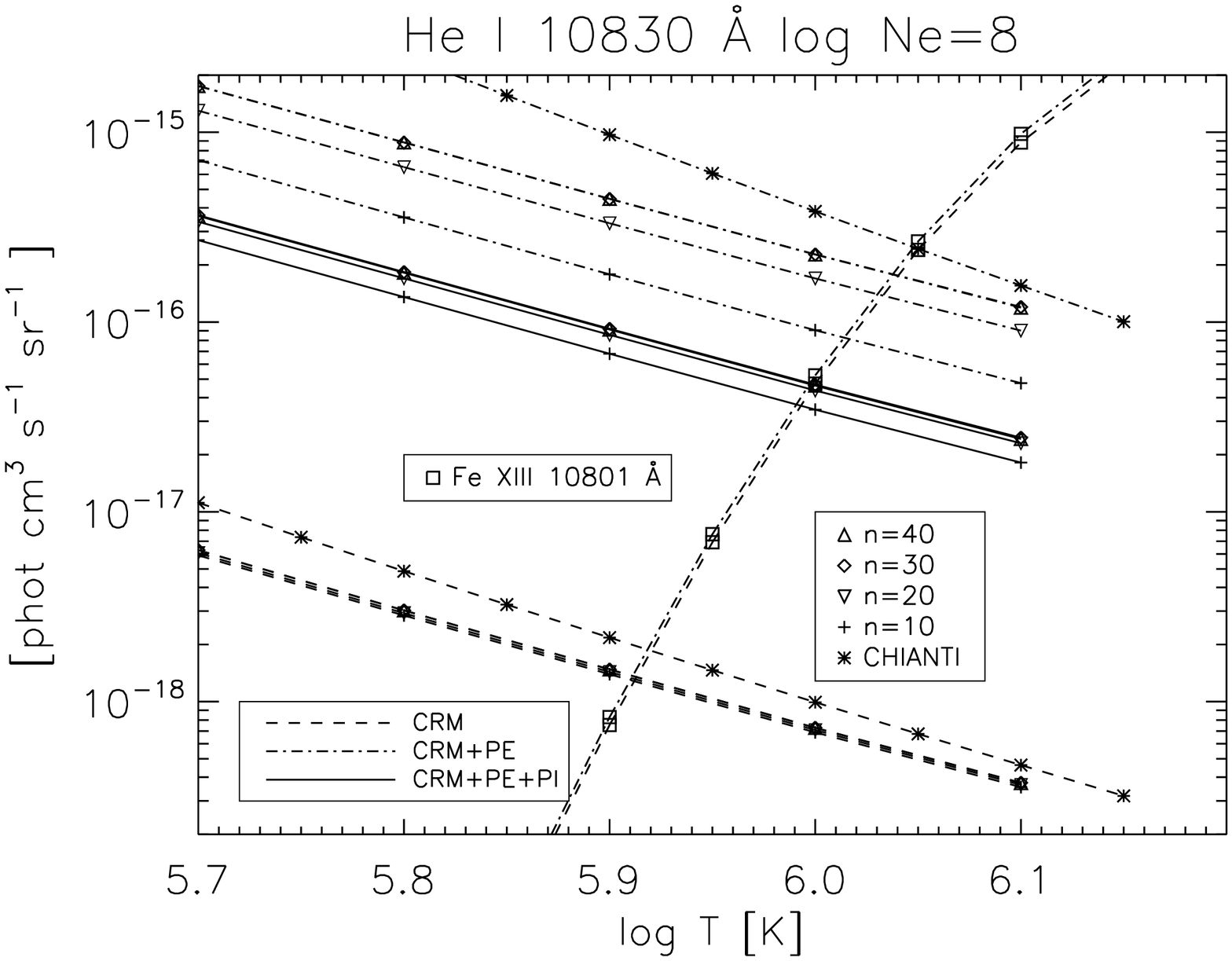}
\includegraphics[angle=-90,width=9.cm]{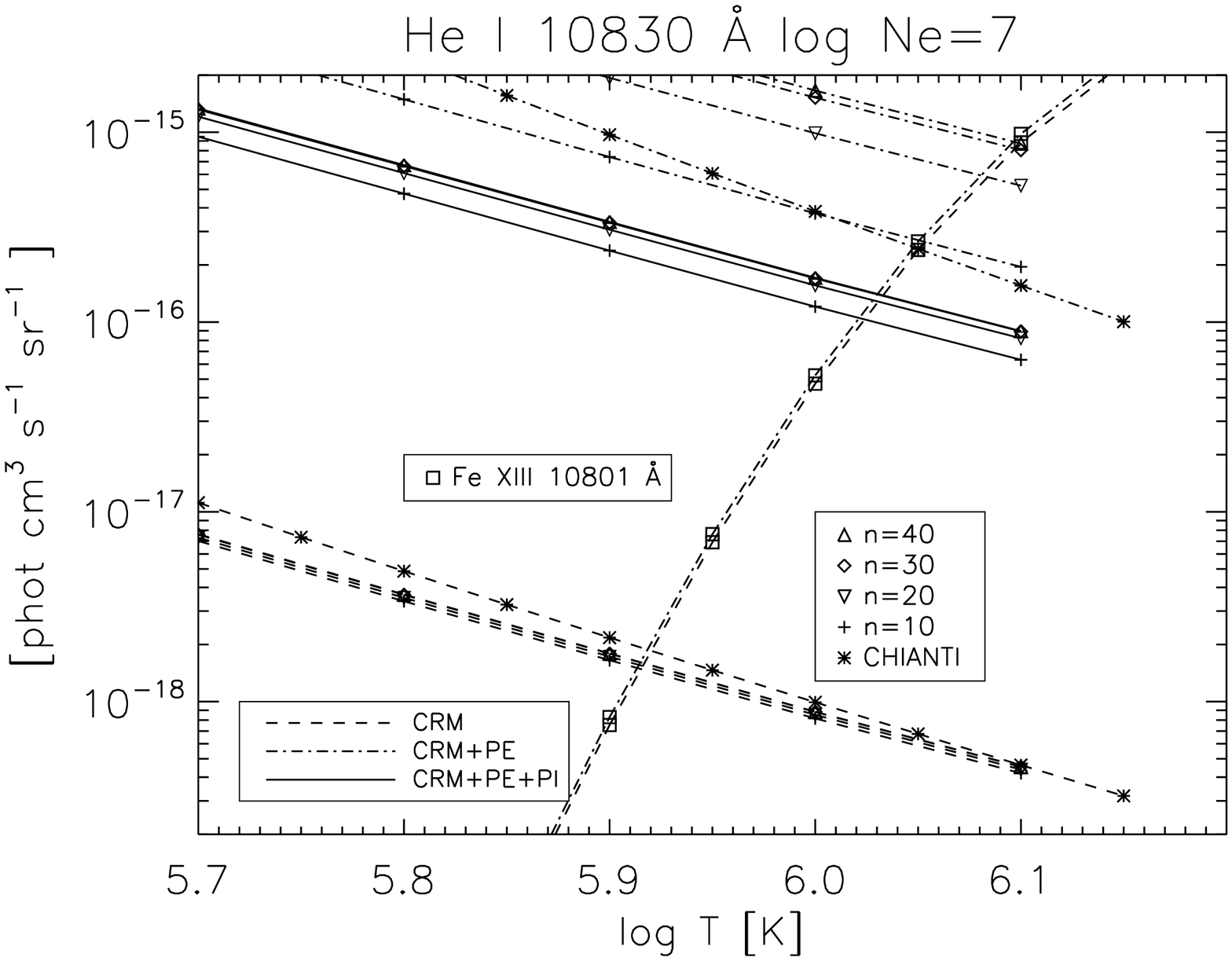}
\end{center}
\caption{Emissivities of the 10830~\AA\ line
obtained with the four collisional-radiative models,
without photo-excitation (PE) and photo-ionization (PI) (dashed lines), with PE (dot-dashed lines) and with both PE and PI (full lines). The emissivities calculated with the CHIANTI He model are also shown, as well as those of the \ion{Fe}{13} NIR line (with and without PE - PI is not affecting this ion). The top plot shows the values at a density for the solar quiet corona near the limb, and with 
 PE/PI at 1.05~\rsun. The lower plot shows the values at a lower density of the outer corona, and with PE/PI at 1.5\rsun.  
}
\label{fig:crm6_he_1_10830}
\end{figure}

The \ion{He}{1} 10830~\AA\ line is very close in wavelength to the two well-known forbidden lines from 
\ion{Fe}{13}, the  3s$^{2}$ 3p$^{2}$ $^3$P$_{1}$-3s$^{2}$ 3p$^{2}$ $^3$P$_{2}$
at 10798~\AA\ and the 
3s$^{2}$ 3p$^{2}$ $^3$P$_{0}$-3s$^{2}$ 3p$^{2}$ $^3$P$_{1}$
at 10747~\AA\ (all wavelengths in air). These three lines are a primary objective for the
coronal DKIST observations.

The eclipse observations by \cite{kuhn_etal:1996} of a
quiescent streamer clearly showed
that close to the solar limb the three lines have comparable intensities.
 With increasing radial distance
from the Sun, the \ion{Fe}{13} 10798~\AA\ decreases its intensity significantly,
while the two other lines show a much slower decrease.
From the \ion{Fe}{13} modelling we know that the 
10798~\AA\ is relatively insensitive to PE, while the 
10747~\AA\ becomes strongly photo-excited with decreasing electron densities.

As the observed radiances are the product of the local emissivities
and the local densities integrated along the line of sight (alos),
comparisons with observations are strongly model-dependent. We defer such modelling to a future paper.

For the present paper, we have  chosen to present a comparison of the
local emissivities of the He lines with that of the \ion{Fe}{13} 10798~\AA\ line
for two typical distances from the Sun which are relevant for
DKIST: 1.05 and 1.5~\rsun\ from Sun centre.
At these distances, our  estimated  local densities 
in quiescent streamers are approximately 10$^8$ and 10$^7$ cm$^{-3}$ respectively.

On a side (but important) note, we would like to point out that
the emissivities of the \ion{Fe}{13} forbidden lines can be used as a reference as the atomic model,
based on the latest calculations by \cite{delzanna_storey:12_fe_13} and available
since CHIANTI version 8 \citep{delzanna_chianti_v8}, is relatively accurate.
In fact, their intensities have not changed significantly, compared to
models based on the earlier (smaller) calculations by \cite{storey_zeippen:2010}.
These lines are strongly affected by cascading effects
from higher levels. Hence, proper atomic modelling typically requires 
a few hundreds of levels and very accurate rates.
This is discussed in \cite{delzanna_mason:2018}. PE effects are significant, but can be calculated accurately. PI effects are negligible, unless there is a solar flare. 

By contrast, PI effects, as discussed further below, can be important in determining the \ion{He}{1} level populations and, in particular, the 10830~\AA\ line emissivity.

More generally, useful insight into the main processes determining the \ion{He}{1} level populations can be obtained by inspecting the most important net rate brackets between the levels, $n_i P_{ij} - n_j P_{ji}$, where $P_{ij}$ and $P_{ji}$ are either the radiative or collisional rates between levels $i$ and $j$.  As an example, Figure~\ref{fig:he_net_rates} displays the main net radiative and collisional brackets at $n_\mathrm{e}=10^8$~cm$^{-3}$ and $T=10^6$~K, at 1.05~\rsun. For clarity of display, only rates between levels with $n\le 5$ are shown in detail.  Levels with $5 < n \le 40$ are grouped in two ``average levels'' (shown as gray blocks), one for singlets and another triplets, while levels with $n>40$ are shown as a single average level.

To highlight the role of PE and PI, the left-hand panel shows the main terms in the statistical equilibrium equations in the purely collisional case.  In this case, the excited levels are populated mainly by radiative cascades from He$^+$ following collisional ionization from the ground level.

As discussed more quantitatively below, when PE and PI are taken into account (right-hand panel), the populations of the triplets become dominated by radiative processes due to the high level of illumination by optical radiation (UV, visible and IR) from the solar disk.  Thus, the metastable level $1s2s ^3S$ becomes crucial in determining the relative populations of the triplet levels.  In particular, the relative populations of levels connected via permitted transitions to level $1s2s ^3S$ are ``locked'' to values dictated by the PE radiation field, nearly independently on temperature and density.

The emissivities for both \ion{He}{1} and \ion{Fe}{13} lines are calculated with a grid of electron temperatures. As a guideline, we expect the local temperature close to the Sun to be around 1 MK, increasing with height to a value around 1.4 MK. Such a value was obtained from SoHO UVCS 1999 observations of two coronal
lines in the 1.4--3~\rsun\ range \citep{delzanna_etal:2018_cosie}. A similar temperature was also obtained from NIR observations of forbidden lines by AIR-Spec during the 2017 eclipse \citep{madsen_etal:2019}. 

We note that, as the distance, density, and temperature
are fixed, and as the PE and PI radiances from the solar disk
are well known, the results shown here basically only depend on the
atomic rates included in the models.
For \ion{Fe}{13}, we adopted the version 8 CHIANTI model and ion
charge states.
We adopted photospheric abundances, as there is now converging
evidence that they better represent the quiescent solar corona
\citep[see the recent review in ][]{delzanna_mason:2018}.
However, as already known for other elements,
there seems to be a significant spatial variability in the coronal He abundance, as discussed in \cite{moses_etal:2020} and references therein.  
Therefore, the present
emissivities should be taken as indicative.

In order to discuss in more detail how the different processes affect the line
emissivities, we present first in Figure~\ref{fig:crm6_he_1_10830} (top)
the  emissivities of the He 10830~\AA\ line, obtained with the four different CRM:
$n=10, 20, 30, 40$ for a density of 10$^8$ cm$^{-3}$ (the lower dashed lines). We then show with the dot-dashed lines the same results obtained with PE, and with full lines those with PE and PI.  
Considering first the case without PE and PI, 
it is clear that the different models show similar results. 
It is interesting to note that, despite the numerous differences in the rates, the emissivities calculated with the present CHIANTI model are not far from our calculated ones.
The same holds for the other He lines as shown below.
It is also clear that the intensity of the He 10830~\AA\ line
is  about two orders of magnitude lower than that of the
\ion{Fe}{13} 10798~\AA\ line at 1 MK.

At 1 MK, the population of the 2p $^3$P, upper level of the He 10830~\AA\ transition, is due by 85\% to collisional excitation
from the metastable 2s $^3$S, the rest from cascading mostly from 
higher $^3$S and $^3$D. In turn, they are populated by further cascading, DR from He$^+$, and collisional excitation from the metastable 2s $^3$S. 
The relative population (over the whole He) of the 2s $^3$S level is 
9.1 $\times$ 10$^{-11}$, that of the ground state is 3.9 $\times$ 10$^{-10}$ and that of the 2p $^3$P  is 1.1 $\times$ 10$^{-15}$.
The emissivity of the He 10830~\AA\ line is directly related to the total DR rate to the triplets from the ground state of He$^+$, hence to the population of this state. In turn, the ground state of He$^+$ is relatively well constrained by the CI from (and DR to) He, plus CI (and RR from) to He$^{++}$.

The results obtained when including PE (dot-dashed lines) are completely different. 
The first main result to note is the large increase in the emissivity
of the He 10830~\AA\ line, by nearly three orders of magnitude. The second result is the fact that the 
$n=10,20$ models are insufficient
to properly estimate the He 10830~\AA\ line.
The $n=30$ model appears sufficient, as the more complex $n=40$ model basically
provides the same values.

The reason for this emissivity increase is the cumulative effect of PE
over many transitions, with consequent cascading to increase by orders
of magnitude the population of the $^3$P, the upper level of the He 10830~\AA\ line.
When PE is switched on, all the levels
that are connected to the metastable 2s $^3$S via transitions in the visible
   and near infrared are photo-pumped by the large number of photons
radiating from the disk. The population of the metastable drops to 
2.9 $\times$ 10$^{-12}$ (that of the ground state is 
4.1 $\times$ 10$^{-10}$) and the 
upper level 2p $^3$P is now populated by 93\%  by the 10830~\AA\ photons, with a relative population of 3.3 $\times$ 10$^{-13}$. 
The remaining population 
   is mainly via cascading from the 3d $^3$D (5877~\AA),
   and 3s $^3$S (7067~AA), so also these
   levels are photo-pumped. With PE, the 2p $^3$P also increases the
   populations of the 4s  $^3$S (4714~\AA), 4d $^3$D (4472~\AA),
5s $^3$S (4121~\AA), and  5d $^3$D (4027~\AA), so in practice all 
the triplets have increased populations (see again Fig.~\ref{fig:he_net_rates}.

\begin{figure}[htbp!]
\begin{center}
\includegraphics[angle=-90, width=8.0cm]{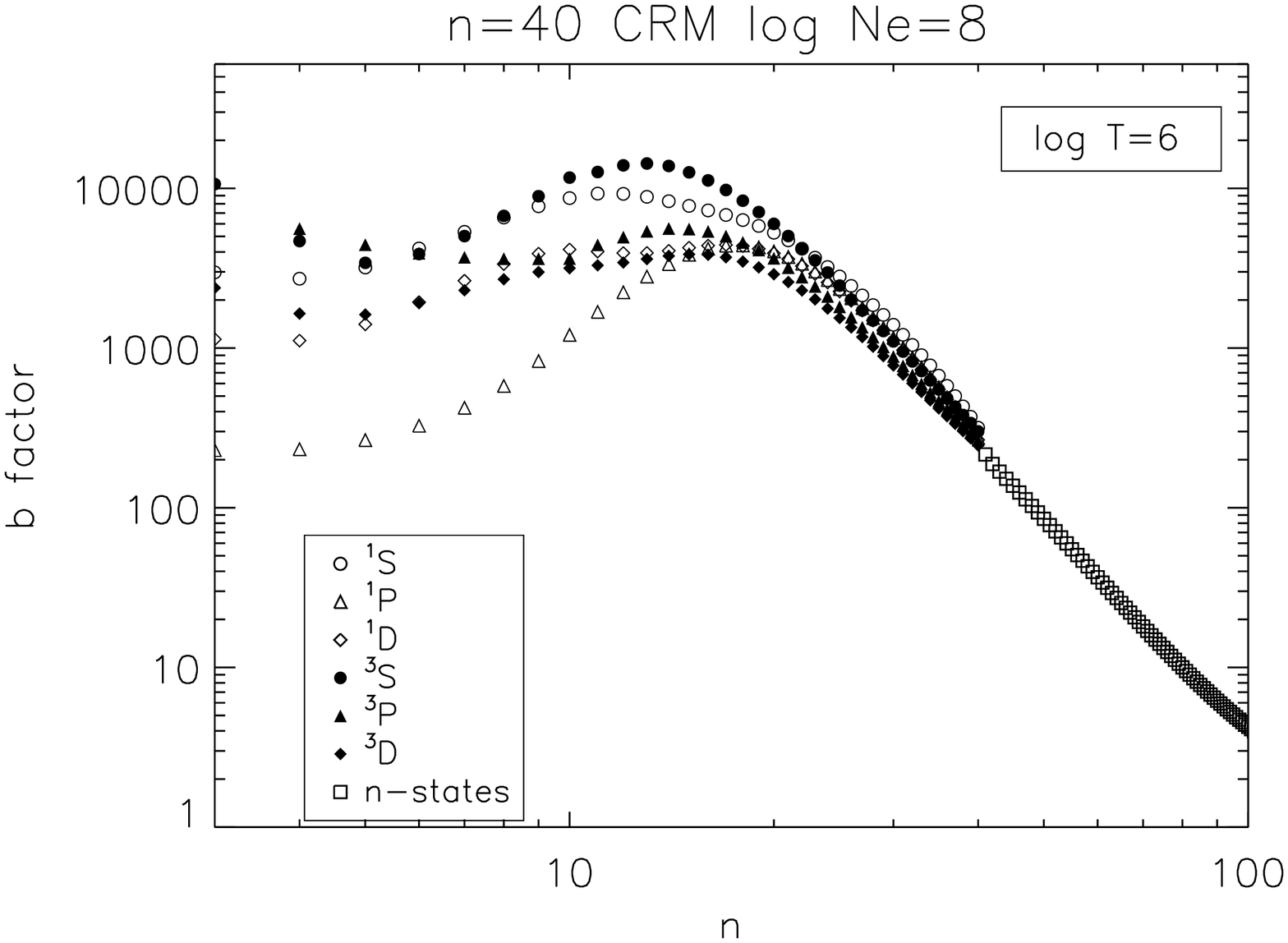}
\includegraphics[angle=-90, width=8.0cm]{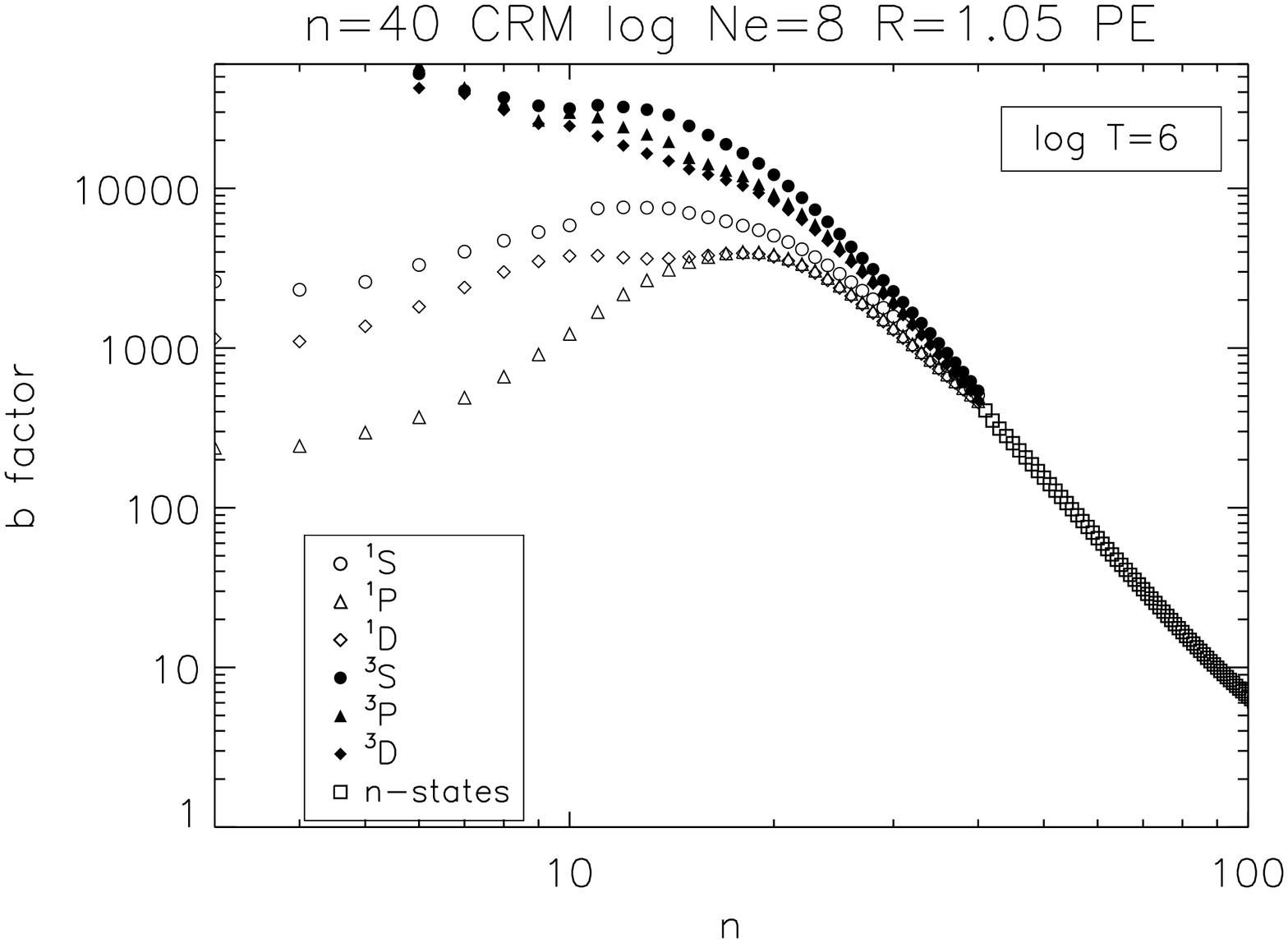}
\includegraphics[angle=-90, width=8.0cm]{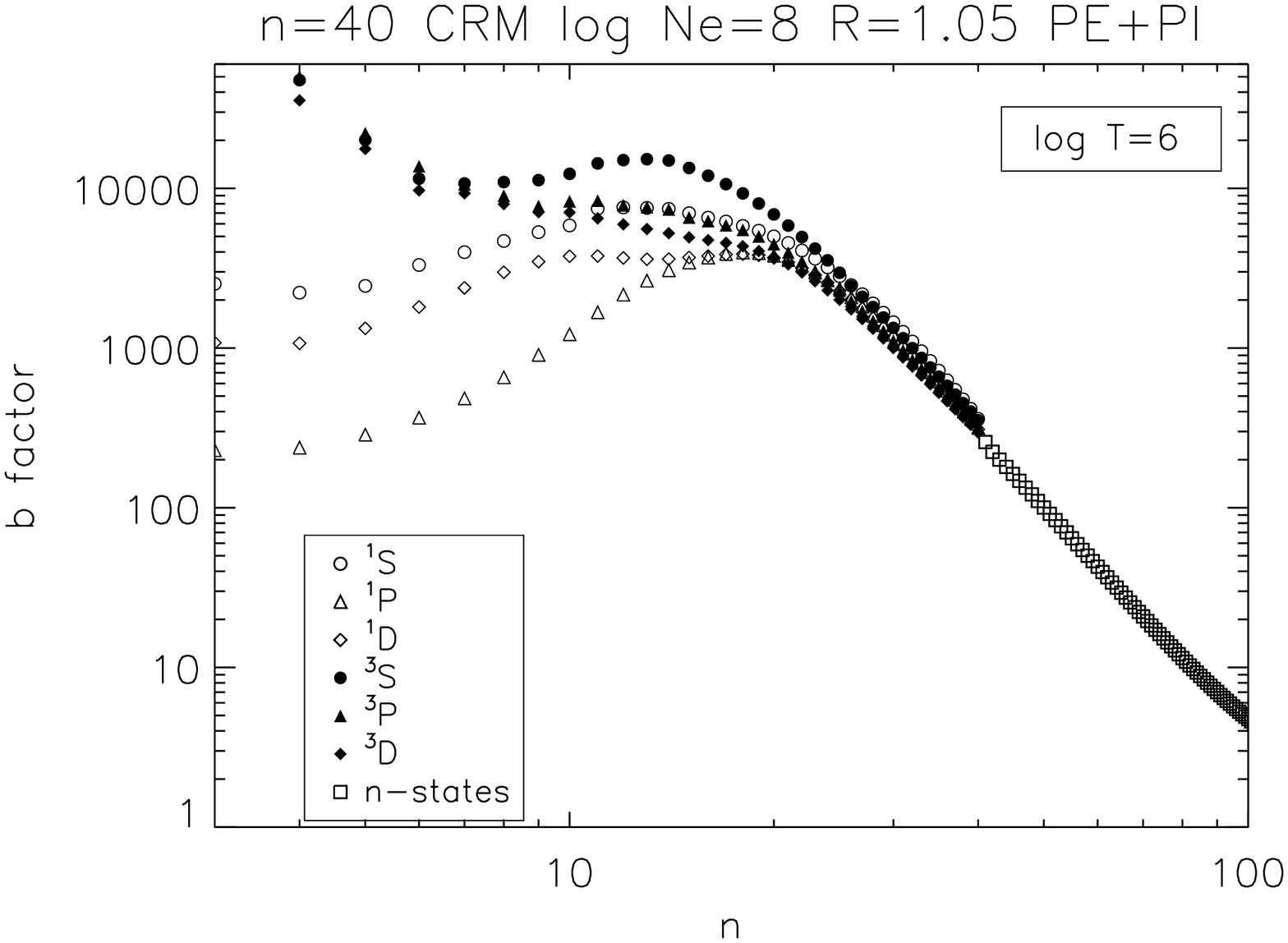}
\caption{b-factors for the solar corona near the limb, without (above),  with (middle) photo-excitation (PE), and with both 
PE and photo-ionisation (PI), below, obtained with the larger $n=40$ model.}
\end{center}
\label{fig:ne8_bfactor}
\end{figure}

One synthetic way to show this effect is to plot the
b-factors, i.e. the deviations from the Saha-Boltzmann population
(relative to the ground state of He$^+$),
as shown in Figure~\ref{fig:ne8_bfactor}.
The top plot shows the b-factors without PE. Note that the 
relative populations of the singlets and triplets converge to
the values of the $n$-levels. At sufficiently high principal quantum numbers, the $n$-levels converge to a Saha-Boltzmann population.
The middle plot shows the b-factors with PE included, where it is 
clear that all the triplet levels have increased populations.


 We also noted that some differences are obtained when the observed spectrum is used instead of a black-body. Long-wavelength photons in the IR above 2.5 microns also play a role, while those below 2000~\AA\ are  insignificant. For more accurate modelling, it would be useful to have observations of the infrared continuum. 
Solar variability should not have a large effect, as observations of the visible/NIR spectrum show very little changes with the solar cycle.

The addition of PI  dramatically affects the population of the metastable He triplet level, which drops to 
5.9 $\times$ 10$^{-13}$ (that of the ground state only has a small decrease to  3.8 $\times$ 10$^{-10}$). As a consequence, 
the upper level 2p $^3$P of the 10830~\AA\  has a much lower relative population of 6.7 $\times$ 10$^{-14}$, and the emissivity of the line therefore decreases accordingly.  
Still, the emissivity of the He line is comparable to that of the reference \ion{Fe}{13} line, around 1 MK. 
The decrease in the population of the triplets is clear in 
the b-factors, shown in Figure~\ref{fig:ne8_bfactor} (bottom).

If only PI and not PE was present, the main effect would be photoionisation of the two 2s $^{1,3}$S levels. However, when both PE and PI are included in the model, photoionisation of the lower excited levels  has a cumulative effect lowering the 2p $^3$P population by a factor of about two.
The main reason this PI is so effective is the fact that the thresholds   are close to the peak solar disk emission in the visible. For example, the PI edge of the 
2s $^3$S is around 2600~\AA (in the MUV), where a significant amount of photons are emitted by the solar surface (still significantly less than assuming a black body).

The solar composite spectrum of \cite{woods_etal:2009_whi} 
has for the MUV wavelengths data from the SORCE Solar Stellar Irradiance Comparison Experiment (SOLSTICE) MUV channel.
\cite{woods_etal:2009_whi} noted a difference of about 10\% 
compared to earlier ATLAS-3 measurements. This is larger than the quoted uncertainties in the measurements. 
Regarding solar variability, it is well known that variations around these wavelengths are not as large as in the FUV/EUV. For example, \cite{rottman:2000} shows that cycle variations at wavelengths shortward of 2600~\AA\ are at most a few percent. Variations and uncertainties in the visible are even less.

The emissivities  of the 10830~\AA\ line at 1.5~\rsun, calculated without PE,
are shown in Figure~\ref{fig:crm6_he_1_10830} (below).
The results are somewhat similar to the previous ones near the limb.
As one would expect, the emissivities obtained when including PE are even more pronounced, as in this case the
density is much lower, 10$^7$ cm$^{-3}$. A lower local density increases the PE effects relative to collisional processes.
It should be noted that this time even the $n=30$ model did not appear
to have reached full convergence, as the much larger $n=40$ model
provides slightly higher  emissivities.
The final case with PE and PI is also similar to the
previous case, although the emissivity of the 10830~\AA\ line is now higher. 
There is clearly a strong temperature
sensitivity: at 1 MK the 10830~\AA\ line emits a lot more photons than the \ion{Fe}{13} 10798~\AA\ line, but at 1.3 MK the He line should be much weaker by over an order of magnitude.

\subsection{\ion{He}{1} \Dthree\ and 584~\AA\ lines}

\begin{figure}[htbp!]
\begin{center}
\includegraphics[angle=-90,width=9.cm]{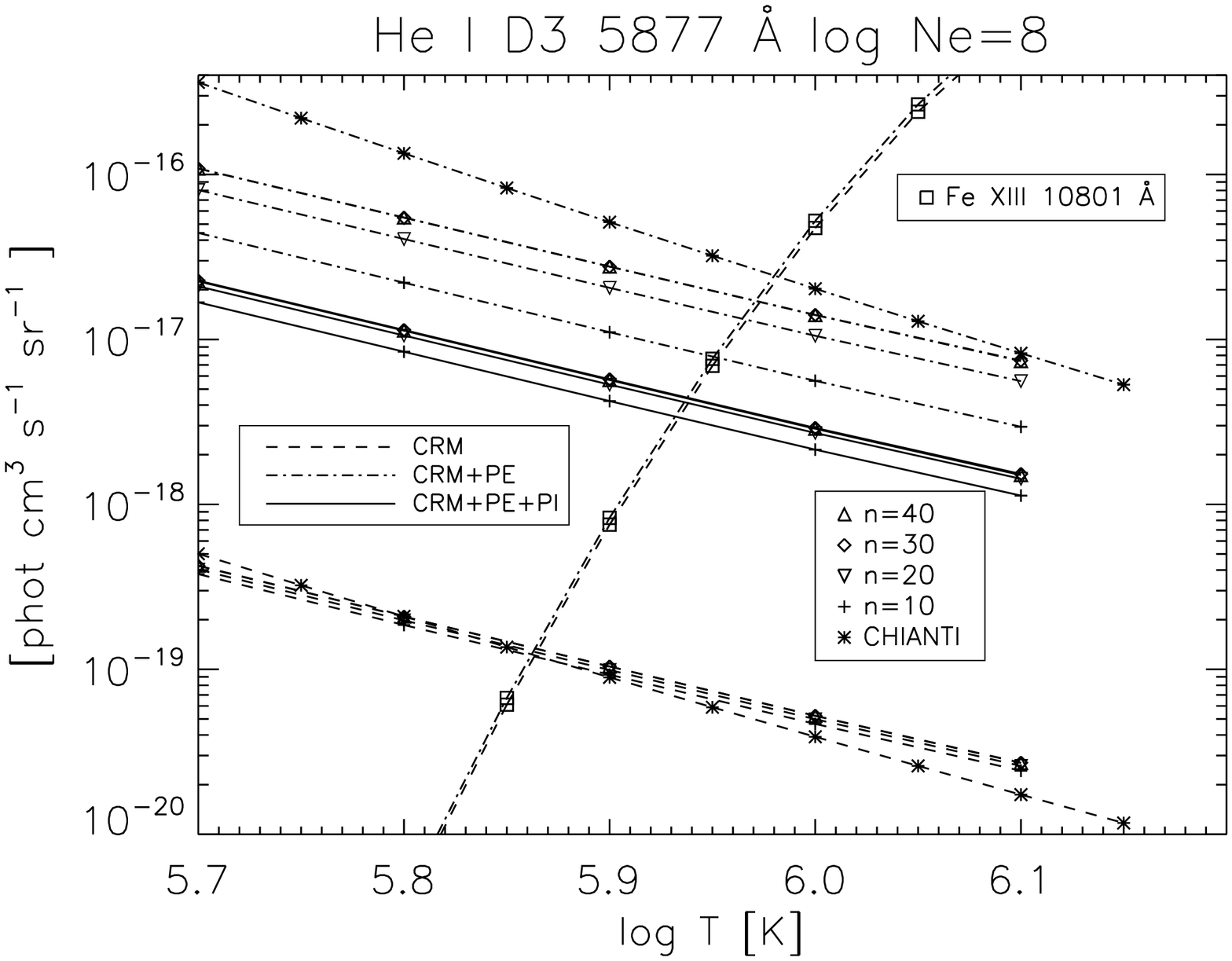}
\includegraphics[angle=-90,width=9.cm]{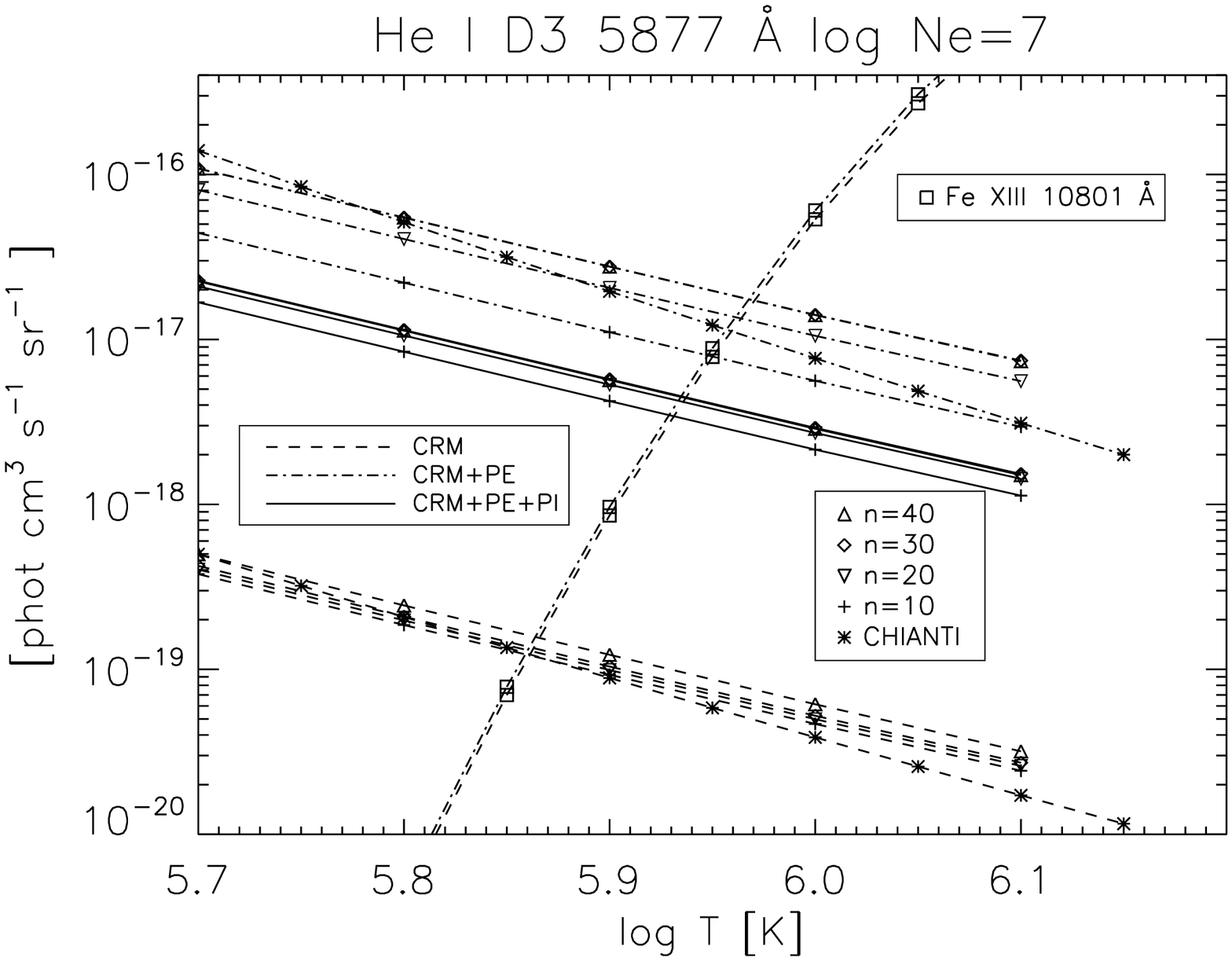}
\end{center}
\caption{Emissivities of the \ion{He}{1} \Dthree\ line
obtained with the four collisional-radiative models,
without photo-excitation (PE) and photo-ionization (PI) (dashed lines), with PE (dot-dashed lines) and with both PE and PI (full lines). The emissivities calculated with the CHIANTI He model are also shown, as well as those of the \ion{Fe}{13} NIR line, with and without PE (PI is not affecting this ion). The top plot shows the values at a density for the solar quiet corona near the limb, and with  PE/PI at 1.05~\rsun. The lower plot shows the values at a lower density of the outer corona, and with PE/PI at 1.5\rsun.  Note the different scale compared to the 10830~\AA\ plot.
}
\label{fig:crm6_he_1_d3}
\end{figure}

Figure~\ref{fig:crm6_he_1_d3} shows the emissivities of the \Dthree\ line, presented in the same way as those of the 10830~\AA\ transition (but on a different scale).
Being a triplet transition, this line is affected by the same processes as the 10830~\AA\ one, and the overall results are very similar. The $n$=30 model shows convergence, and the same patterns when PE and PI are present.
The emissivity of the \Dthree\ line, however, even at 1 MK is predicted to be over one order of magnitude weaker than the \ion{Fe}{13} line, hence it might not be visible by ASPIICS outside of prominences.

\begin{figure}[htbp!]
\begin{center}
\includegraphics[angle=-90,width=9.cm]{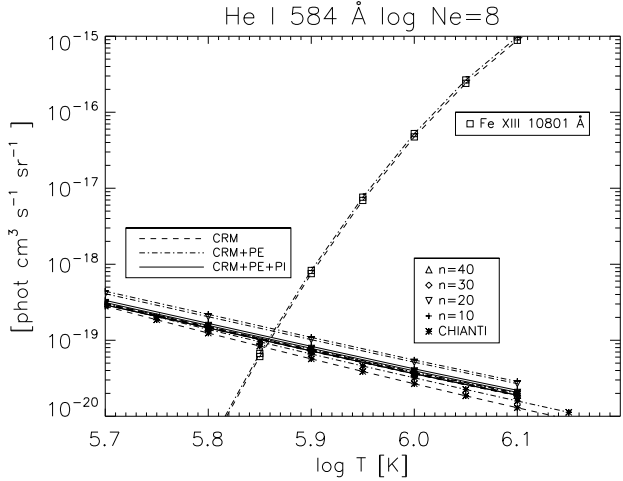}
\end{center}
\caption{Emissivities of the He 584~\AA\ resonance line
obtained with the four collisional-radiative models, displayed as in 
Figure~\ref{fig:crm6_he_1_d3} for the solar quiet corona near the limb, and with  PE/PI at 1.05~\rsun.
}
\label{fig:crm6_he_1_584}
\end{figure}

Figure~\ref{fig:crm6_he_1_584} shows the results for the 
 resonance line at 584~\AA,
the decay from the $^1$P directly to the ground state,
for the solar quiet corona near the limb, and with  PE/PI at 1.05~\rsun.
As there are comparatively
very few photo-exciting photons originating from the disk,
PE hardly has any effect on this line (and on any other lines of the principal series).
We therefore predict that this line should be about 
three orders of magnitude weaker than the \ion{Fe}{13} NIR  line, i.e. practically invisible. 
Our findings are consistent with the lack of observations of that line by SoHO UVCS in the near-Sun corona that could be unambiguously be attributed to coronal plasma (see Sec.~\ref{sec:intro}).

\subsection{Ionised helium}

\begin{figure}[htbp!]
\begin{center}
\includegraphics[angle=-90,width=8.0cm]{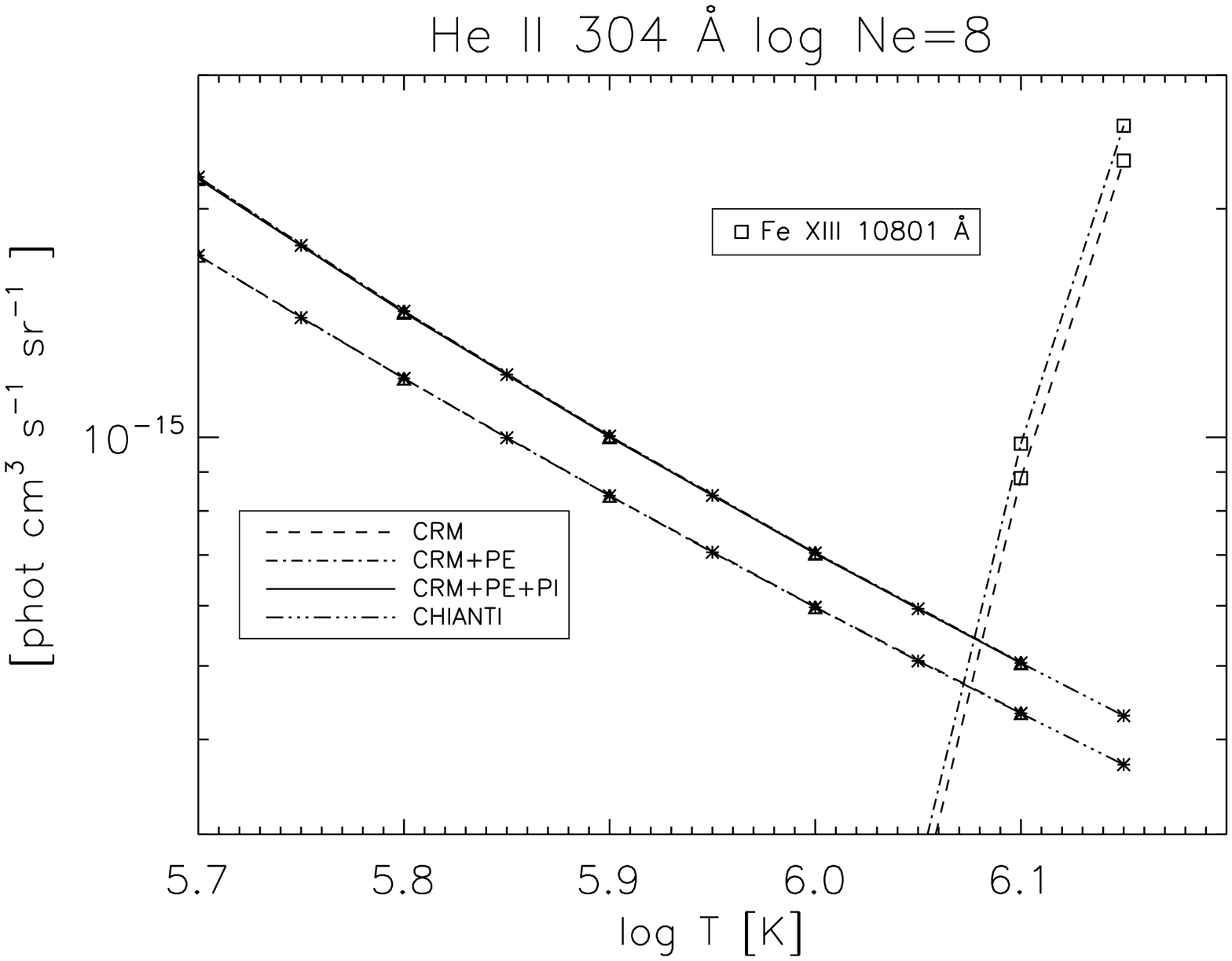}
\includegraphics[angle=-90,width=8.0cm]{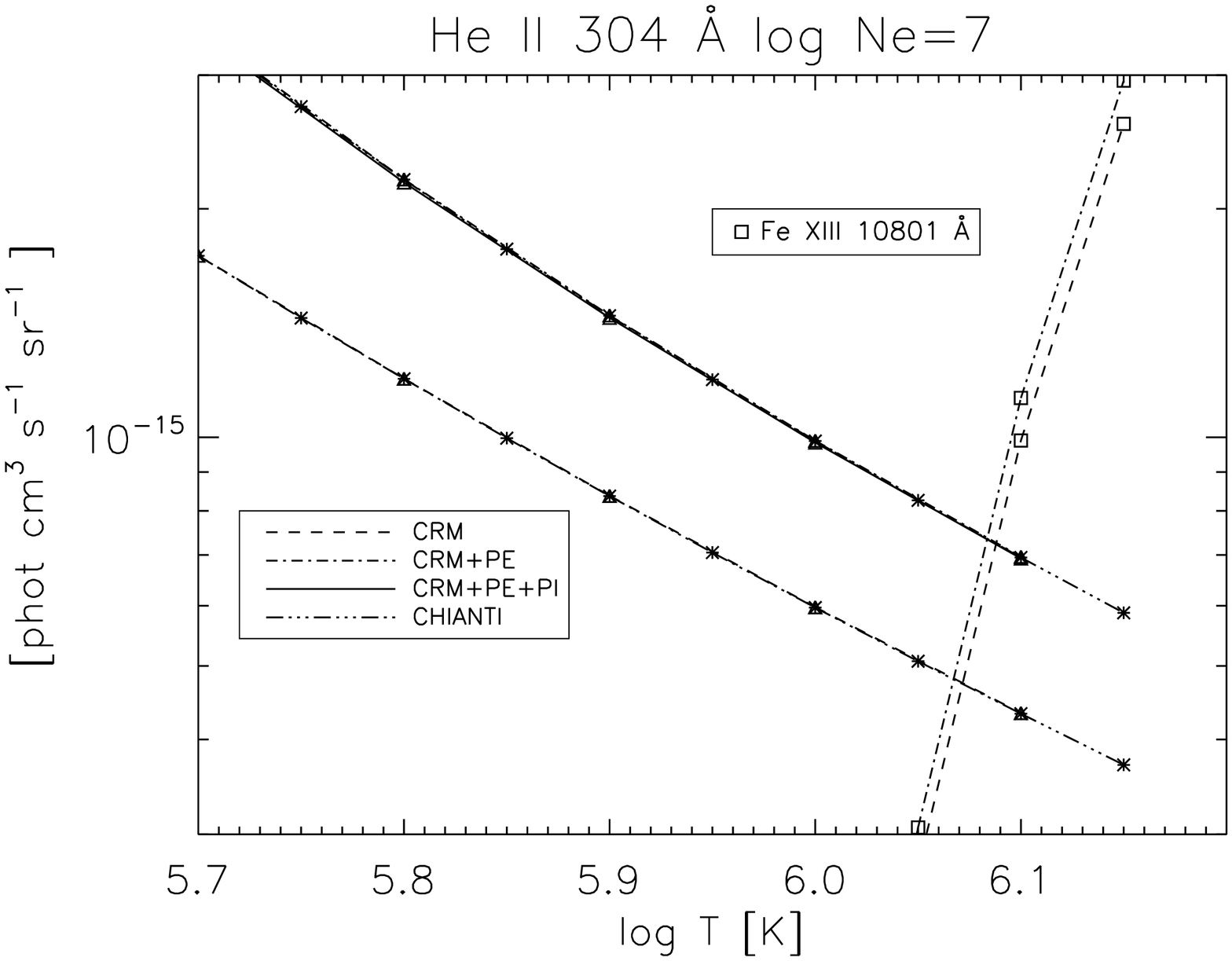}
\caption{Emissivities of the main He$^+$ resonance line at 304~\AA\
for the solar corona near the limb at 1.05~\rsun\ (top) 
and at 1.5\rsun\ without and with 
photo-excitation (PE) and photo-ionisation (PI), obtained with the largest $n=40$ collisional-radiative model.
The emissivity of the \ion{Fe}{13} NIR line with (and without) PE
is shown with a full (dashed) line for comparison. 
Note the different scale compared to the the 10830~\AA\ plot.}
\end{center}
\label{fig:crm6_he_2}
\end{figure}

The modelled emissivities of the main He$^+$ resonance line are shown in Figure~\ref{fig:crm6_he_2} in the same way as those of the 10830~\AA\ transition (but on a different scale).
This line is strongly affected by photo-excitation (PE), especially at low electron densities. Therefore, direct measurements of the variable disk radiation will be needed for detailed models.
On the other hand,
 its emissivity is rather insensitive to the model used or to photo-ionisation (PI) for coronal temperatures.  The emissivities are close to those obtained
with the CHIANTI model (modified to input the same resonant PE), mainly because in both cases the same ionization and recombination rates to He$^{++}$have been used, and because the same He$^+$ ion model is adopted.

\section{Conclusions}

As a result of a careful assessment of all the rates and the effects of photo-excitation and photo-ionisation on progressively more extensive collisional-radiative models, we are able to recommend a model to be used to study the helium emission in the solar corona. The key issue for the neutral He emission is that a fully $LS$-resolved set of states up to $n=40$, and $n$-resolved states up to $n=300$
are required. The complexity of this model is similar to the most extensive models developed to study He for photo-ionised gas in nebulae, but in very different conditions, i.e. much lower temperatures and densities than in the coronal case. The revised dielectronic recombination rates presented here resulted in a significant (factor of two) increase in the abundance of neutral He.

Regarding He$^+$, we confirm that the resonance 304~\AA\ line is a strong coronal line, mainly because at coronal temperatures He$^+$ has a significant abundance. This line also has a strong resonantly scattered component which increases its intensity. Therefore, solar variability plays an important role, as the disk radiation is variable. As the intensity of this line is directly related to the population of the ground state of  He$^+$, which in turn affects the population of the He levels by DR, an ideal observational benchmark of the present model would include simultaneous observations of the He and He$^+$ lines. Indirect measurements will be possible with a combination of Solar Orbiter and DKIST observations. 

Regarding neutral He, 
as described in the introduction, several possible explanations
have been put forward to explain the apparently anomalous high intensity of the 10830~\AA\ line in the outer corona. 
We note that in general, various mechanisms could be at play.
The purpose of the present paper was not to discuss them, but provide a key missing element: a prediction of the coronal emission of the main \ion{He}{1} lines, in the light of the  new polarimetric measurements  
by DKIST and Proba-3 which in principle will provide a novel way to measure the magnetic field in the low corona.

The present models indicate that a significant coronal
emission of the 10830~\AA\ line should be present
close to the solar limb, for an electron temperature of 1 MK. In this case, the intensity of this line should be comparable to that of the
well-studied \ion{Fe}{13} 10798~\AA\ line, one
of the strongest forbidden lines in the solar corona
\citep{delzanna_deluca:2018}. 
This appears in line with observations and is mainly due to a complex system of photo-excitation (PE) and photo-ionisation (PI) due to the disk
radiation, with subsequent cascading (and recombination followed by cascading) to over-populate the upper level of this transition. 
The PE and PI effects are sensitive to the solar spectrum. We have briefly mentioned where accurate measurements would be useful. We also expect that some minor solar variability effects would be present.

However, in the outer corona, the emissivity of the 10830~\AA\ line should decrease significantly, with respect to the \ion{Fe}{13} 10798~\AA\ line.  The actual observed emission would mainly be the result of a fine balance between the distributions of electron densities and temperatures along the line of sight. 
Independent measurements of the density and its fine structure with e.g. a combination of the \ion{Fe}{13} NIR and EUV lines would be needed for a proper modelling of the \ion{He}{1} lines, as well as estimates of the electron temperature, which is harder to measure. 
Finally, the coronal abundance variability needs to be taken into account.

The \ion{He}{1} \Dthree\ line is also affected strongly by PE and PI, although this feature is much weaker and might not be observable. 
On the other hand, the EUV resonance 584~\AA\  line is not pumped by 
PE, and should be unobservable. 
Any observations of the 584~\AA\  line in the
outer corona would be important, as they would indicate that
other processes augment the He emission, as
for example the presence of dust or cool gas from e.g. erupting prominences,
as proposed in the literature. 

We plan to apply the present models to provide more detailed predictions of the expected  helium emission in the corona, as will be observed by several new facilities, by taking into account the 
key input parameters and their variability.
We also plan to apply them to
revisit the modelling of the He lines in the transition region, i.e. at lower temperature and higher density conditions.

As the basic rates and models are very general, we also plan to extend them to predict the recombination spectrum of He for nebular conditions. The comparison with literature values at the lowest density achievable by the largest model has clearly indicated that improvements on previous studies can be made.

\acknowledgments
GDZ acknowledges support from STFC (UK) via the consolidated grants 
to the atomic astrophysics group (AAG) at DAMTP, University of Cambridge (ST/P000665/1. and ST/T000481/1).
NRB acknowledges support from STFC (UK) via the UK APAP Network grant (ST/R000743/1)
with the University of Strathclyde.
We would like to express our gratitude to 
our colleagues
in the AAG for stimulating discussions:
N. Ljepojevic, who developed COLRAD to model
H-like ions;  I. Grant, who developed the relativistic atomic
structure theory; R. Dufresne, who tested a simplified version
of the CRM for Carbon;
H.E. Mason, one of the founders of the CHIANTI atomic database. 


We also thank several colleagues who attended the SHINE 2019 Workshop for their encouragement, in particular G. Cauzzi and V. Martinez Pillet who contributed to the organization of the session No. 16, and J. Kuhn and J. Raymond for their insightful reviews of the theme related to the observations of neutrals in the corona. 


\bibliography{helium_crm}{}
\bibliographystyle{aasjournal}



\end{document}